\documentclass[onecolumn,12pt,twoside]{IEEEtran}
\usepackage{bbm}
\usepackage{amsfonts}
\usepackage{tipa}
\usepackage{amssymb}
\usepackage{mathrsfs}
\usepackage[lined,boxed,commentsnumbered]{algorithm2e}
\usepackage{graphicx}
\usepackage{amsfonts,amssymb,amsmath}
\usepackage{latexsym}
\usepackage{multirow}
\usepackage{hyperref}
\usepackage{epsfig,epstopdf}

\newtheorem{rem}{Remark}
\newtheorem{lemma}{Lemma}

\newtheorem{Proposition}{Proposition}

\normalsize

\begin{document}
\title{Weighted Sum-Throughput Maximization for MIMO Broadcast Channel: Energy Harvesting Under System Imperfection}
\author{Zhi~Chen~\IEEEmembership{Member,~IEEE} Pingyi~Fan~\IEEEmembership{Senior Member,~IEEE}
Dapeng~Oliver~Wu~\IEEEmembership{Fellow,~IEEE}
        and~Khaled~Ben~Letaief~\IEEEmembership{Fellow,~IEEE}
\thanks{Z. Chen and P. Fan are with the Department of Electrical Engineering, Tsinghua University, Beijing, China, 100084. Emails: chenzhi2223@gmail.com; fpy@tsinghua.edu.cn.
D. Wu is with Department of Electronic and Computer Engineering, University of Florida, Florida, USA, 32611 (e-mail:wu@ece.ufl.edu).
K. B. Letaief is with Department of Electronic and Computer Engineering,
Hong Kong University of Science and Technology, Hong Kong (e-mail:
eekhaled@ece.ust.hk).}}

%\markboth{Submit to IEEE Transaction in Signal Processing.}

\maketitle

\baselineskip=24pt
\begin{abstract}\\
\baselineskip=18pt
In this work, a MIMO broadcast channel under the energy harvesting (EH) constraint and the peak power constraint is investigated. The transmitter is equipped with a hybrid energy storage system consisting of a perfect super capacitor (SC) and an inefficient battery, where both elements have limited energy storage capacities.
In addition, the effect of data processing circuit power consumption is also addressed.
To be specific, two extreme cases are studied here, where the first assumes ideal/zero circuit power consumption and the second considers a positive constant circuit power consumption where the circuit is always operating at its highest power level. The performance of these two extreme cases hence serve as the upper bound and the lower bound of the system performance in practice, respectively.
In this setting, the offline scheduling with ideal and maximum circuit power consumptions are investigated. The associated optimization problems are formulated and solved in terms of weighted throughput optimization. Further, we extend to a general circuit power consumption model.
To complement this work, some intuitive online policies are presented for all cases. Interestingly, for the case with maximum circuit power consumption, a close-to-optimal online policy is presented and its performance is shown to be comparable to its offline counterpart in the numerical results.
\end{abstract}
\begin{keywords}
MIMO-BC, energy harvesting, battery imperfection, resource allocation.
\end{keywords}
\IEEEpeerreviewmaketitle

\section{Introduction}
In recent decades, MIMO communication has attracted increasing attention due to its capability in improving network capacity without incurring additional bandwidth or power usage. A typical example is the MIMO broadcast (BC) channel, consisted of one multi-antenna transmitter (base station) and several multi-antenna receivers (mobile users), as employed in the cellular downlink transmission.
It was extensively studied in \cite{Shamai06}-\cite{Tran07} on its capacity region, capacity-achievable approach, as well as the optimal power allocation policy. As observed in \cite{Shamai06}\cite{Goldsmith05}\cite{Love07}, dirty paper coding (DPC) was shown to be capable of achieving the capacity region, however the proposed successive water-filling strategy was of high complexity. Therefore, zero-forcing DPC (ZF-DPC) was introduced in \cite{Caire03}\cite{Dabbagh07} to reduce the problem of precoder design by decomposing a MIMO BC into a series of parallel interference-free point-to-point channels, while its performance close to the capacity region was also demonstrated.
To further reduce computation complexity, the authors in \cite{Tran07} proposed a ZF-DPC strategy based on the single QR decomposition (QRD), which is adopted in this work to find the transmit covariances.

In addition, due to the smart exploitation of available energy resources in the ambient environment, energy harvesting (EH) powered communication systems arise in recent decades \cite{Kansal07}-\cite{Ozel14}, where various channel models with EH constraints were studied.
In \cite{Kansal07}-\cite{Gupta2010} energy management policies for EH enabled sensor networks were investigated.
In \cite{Ulukus2012harvesting} and \cite{Ulukus2010harvesting} the optimal off-line scheduling for one-hop transmission over static channels was designed, provided full knowledge of energy harvested and link gains in all considered slots from the beginning of transmission at the transmitter side.
In \cite{Zhang2011allocation}-\cite{Devillers2011harvesting}, the optimal off-line scheduling for relay networks over static channels was investigated.
In \cite{Ulukus2011fading}-\cite{Zhang2012allocation} the optimal one-hop transmission policy over fading channels was studied, for both the off-line scheduling and online scheduling. Specifically, in online scheduling, only causal information of the link gains and harvested energy of the current and previous slots is known at the transmitter.

Compared with the previous works assuming perfect batteries with infinite capacity and no energy storage loss, in
\cite{Ozel12}\cite{Yener2012harvesting}, optimal transmission policies under EH constraints with a finite capacity battery were investigated. In \cite{Gunduz12}, battery imperfection was addressed for a general framework for EH communication optimization. In \cite{Yener12}, energy storage loss was investigated for a one-hop unicast system with EH constraints.
Further, in \cite{Xu12}-\cite{Ozel14}, circuit power consumption as well as transmit power consumption were both taken into account for throughput optimization over a point-to-point link.
More interestingly, in \cite{Ozel14}, a hybrid energy storage system was considered for a point-to-point Gaussian channel, consisting of an imperfect battery with unlimited storage capability and an ideal super capacitor (SC) with limited storage capability. It is worthy to note that the physical peak power constraint was not taken into account in these works.

In fact, the data processing circuit power consumption is an important factor which affects system performance for a communication system with energy harvesting. Hence, we shall take it into account in this work. To do so, two extreme cases are firstly studied, where the first one is that the data processing circuit power consumption is ideal/zero, which is the ideal target of the future circuit design. In fact, we believe that the circuit power consumption will become much lower in the future along with the development of novel circuit design techniques as well as new introduced/discovered circuit materials. The second one is that the data processing circuit power always consumes at its highest level (a constant level for sure), which can be regarded as the circuit system in the transmitter always works with its maximum admitted power. Further, the general case where the circuit power consumption ranges between these two extreme values are discussed to complement this work.

In this work, we are hence motivated to study the weighted sum throughput problem for a MIMO BC channel under EH constraints and peak transmit power constraints, which is commonly employed in cellular networks on the downlink. A hybrid energy storage system of two energy buffers is considered, where both buffers are with limited storage capacity. In addition, the impact of circuit power consumption on system performance is investigated. For offline scheduling, convex optimization tools are employed to find the optimal weighted sum throughput with a given deadline. Some observations are made based on the optimal Karush-Kuhn-Tucker (KKT) conditions.
On the other hand, for online scheduling, we utilize the observations found in solving its offline counterpart to design a good online policy. Interestingly, for the case with imperfect circuit, a close-to-optimal online policy is proposed. The contributions of this work are hence summarized as follows.
\begin{itemize}
\item Under both perfect circuit and nonzero circuit power consumption scenarios, the associated optimization problems to obtain the maximum weighted sum throughput are formulated and solved for offline scheduling, where the sum power allocation is found to be of the water-filling structure with a ceiling level.
\item For optimal offline scheduling, some observations on energy transfer and transmit power allocation through time are derived, based on the KKT conditions.
\item For online scheduling, some heuristic algorithms are designed for both scenarios and are shown to perform good in simulations. Especially, for the case with nonzero circuit power consumption, the proposed online algorithm is shown to perform comparable to its offline counterpart.
\end{itemize}

The rest of this work is structured as follows. In Section II, system model of our work is presented. In Section III and Section IV, the offline weighted sum throughput optimization problems are formulated and solved, for both cases with zero or nonzero circuit power consumption, respectively. Some observations are also made based on the optimal solutions to the associated problems. In Section V, an extension to the general circuit power consumption scenario is discussed. In Section VI, the online policy for all cases are designed. Numerical results are presented in Section VII and we conclude this work in Section VIII.

\section{System Model}
\begin{figure}[t]
   \centering
   \includegraphics[width = 7cm]{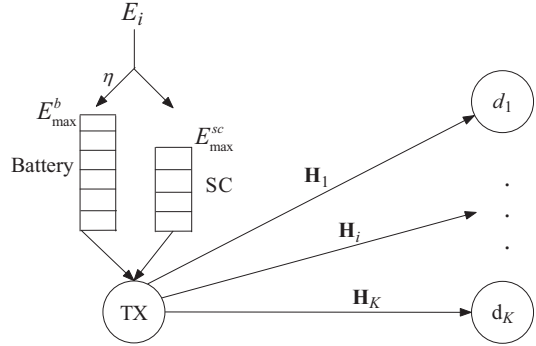}
   \caption{System model of
   a MIMO BC channel with a hybrid energy storage system, where energy storage efficiency is ideal for the SC and $\eta$ for the battery.} \label{fig:sys}
   \end{figure}

In this work, a MIMO broadcast channel consisting of one transmitter/base station and $K$ receivers/users is studied, as shown in Fig. \ref{fig:sys}, where the transmitter is equipped with $M$ ($M>1$) antennas and the $k$th user is equipped with $n_k$ ($n_k \ge 1$ and $k=1,\ldots,K$) antennas. It is also assumed here that $M\ge \sum_{k=1}^K n_k$.
The channel between the transmitter and the $k$th receiver is denoted by a matrix $\mathbf{H}_k \in  \mathcal{C}^{n_k \times M}$, where each entry of the matrix is modeled by a mutually independent Gaussian random variable with zero mean and unit variance. The noise vector at receiver $k$ is denoted by $\mathbf{n}_k$ and is modeled by a complex Gaussian variable vector with zero mean and covariance matrix $\mathbf{I}_{\mathbf{n}_k}$ where $\mathbf{I}_{\mathbf{n}_k}$ is a $n_k \times n_k$ identity matrix.

\begin{figure}[t]
   \centering
   \includegraphics[width = 7cm]{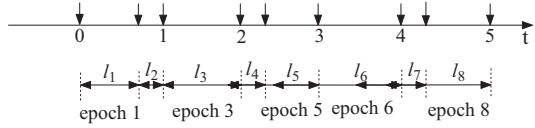}
   \caption{The system model with random energy arrivals, with $7$ random energy arrivals exclusive of the initial energy arrival at the beginning of the transmission. Hence totally there are $8$ epochs in transmission, where the epoch is defined as the duration between two adjacent energy arrival events. The deadline in this example is set to be 5 seconds.} \label{fig:profile}
   \end{figure}

In addition, the transmitter is assumed to have two energy buffers consisting of an inefficient battery and an ideal super-capacitor (SC), where the battery is assumed to have a large storage capacity of $E_{\max}^{b}$ units of energy while the SC can store at most $E_{\max}^{sc}$ ($E_{\max}^{sc}<E_{\max}^{b}$) units of energy. The battery is assumed to be inefficient in energy storage in the sense that the energy drained from it is less than the amount stored. On the contrary, the SC is perfect in energy storage.
It is assumed that $E_i$ amount of energy arrives at $t_i$ at the transmitter.
The transmitter stores $E_i^{sc}$ and $E_i^{b}=E_i-E_i^{sc}$ in the energy buffer of the SC and the battery at time $t_i$, respectively. The initial amounts at the beginning of transmission at the SC and the battery hence are $E_0^{sc}$ and $E_0^{b}$, respectively. Throughout this work, an epoch is defined as the time duration between two energy arrivals, i.e., epoch $i$ is the time interval $[t_i,t_{i+1}]$ and the associated length is $l_i=t_{i+1}-t_i$, as depicted in Fig. \ref{fig:profile}. We also denote a random variable $N$ as the number of energy arrivals by deadline T.
In addition, the energy storage efficiency of battery is assumed to be $\eta$, where only $\eta E_i^b$ can be drained from the battery assuming $E_i^b$ amount of energy is stored at battery. Due to the fact that the SC has perfect storage efficiency, we are tempted to allocate the incoming energy to the SC first and then allocate the rest to the battery.
However, it will be observed that it is beneficial to smooth the
transmit power sequences with the aid of the battery, although at
the cost of some energy loss.
%Hence sometimes we may need to transfer energy from the SC to the battery to smooth the transmit power sequences.
%We denote $\delta_i$ as the energy transferred from SC to battery in epoch $i$ and hence pnly $\eta\delta_i$ can be drained form the battery due to the storage inefficiency at battery.
%It is also assumed that this transferred energy during epoch $i$ becomes available for use in epoch $i+1$.
In addition, it is noted that the sum transmit power $p_i$ at epoch $i$ is $p_i=p_i^{sc}+p_i^{b}$ where $p_i^{sc}l_i$ and $p_i^{b}l_i$ are drained from the SC and the battery, respectively.

Note that our goal is to maximize the weighted sum throughput before the deadline, where for simplicity a natural ordering of users is assumed in this work. However, it is worth noting that different ordering can affect the achieved sum throughput, as observed in the literature.
In this work, zero-forcing DPC is employed to decompose
the MIMO broadcast channel into $K$ parallel SISO Gaussian channels \cite{Caire03}\cite{Dabbagh07}\cite{Tran07}.
To be specific, for receiver $k$, the interference caused by receiver $1$ to receiver $k-1$ is canceled by DPC, and the interference caused by receiver $k+1$ to receiver $K$ is eliminated by zero forcing at each epoch, i.e., by designing that $\mathbf{H}_k\mathbf{W}_j(i)=0$ for $j>k$ where $\mathbf{W}_j\in \mathcal{C}^{M \times n_j}$ is the precoder of the $j$th receiver at epoch $i$.
Hence, the resulting rate of the $k$th user at epoch $i$ is given by
\begin{align}
R_k^{ZF-SPC}(i)=\log|\mathbf{I}+\mathbf{H}_k\mathbf{W}_k(i)\mathbf{W}_k^H(i)\mathbf{H}_k^H|. \label{eq:zf_DPC_user_k}
\end{align}
Further, we can write $\mathbf{W}_k(i)=\mathbf{B}_k(i)\mathbf{D}_k(i)$ where
$\mathbf{B}_k(i) \in \mathcal{C}^{M \times \bar{n}_k}$ is designed to remove the interference where $\bar{n}_k=M-\sum_{i=1}^{k-1}n_k$, and
$\mathbf{D}_k(i)\in \mathcal{C}^{\bar{n}_k \times n_k}$ is for performance optimization.
It is hence required $M > \sum_{i=1}^{K}n_k $ for all users to guarantee $\mathrm{rank}(\mathbf{B}_k(i))=\bar{n}_k \ge >\mathrm{rank}(\mathbf{H}_k(i))=n_k$ ($\forall k$).
It is further observed that by design $\mathbf{B}_k(i)$ should lie in the null space of $\mathbf{\tilde{H}}_k$, i.e., $\mathcal{N}(H_k)$, where $$\mathbf{\tilde{H}}_k=[\mathbf{H}_1^T\mathbf{H}_2^T\ldots
\mathbf{H}_{k-1}^T] \in \mathcal{C}^{\sum_{i=1}^{k-1}n_i \times N}.$$
$\mathbf{B}_k(i)$ is chosen to be an orthonomal basis of $\mathcal{N}(H_k)$ in \cite{Dabbagh07} where the singular value decomposition (SVD) was utilized to find the optimal precoder at a relatively high complexity.
However, noting that
$\mathrm{rank}(\mathbf{H}_k\mathbf{B}_k)=n_k \leq \mathrm{rank}(B_k)=M-\bar{n}_k$ by the assumption $M \ge \sum_{k=1}^K n_k$, $\mathbf{B}_k(i)$ need not be a basis for $\mathbf{\tilde{H}}_k$, instead, it is valid in removing interference as long as it is in the subspace of $\mathcal{N}(H_k)$. In this sense from \cite{Tran07}, we have
$$\mathbf{H}_k\mathbf{W}_k=\mathbf{H}_k\mathbf{B}_k\mathbf{D}_k
=[\mathbf{L}_k, \, \mathbf{0}] [\mathbf{D}_{k,1}^T,\,\mathbf{D}_{k,2}^T]^T=\mathbf{L}_k\mathbf{D}_{k,1}.$$
where $\mathbf{0}$ follows from $\mathrm{rank}(\mathbf{H}_k)<\mathrm{rank}(\mathbf{B}_k)$. $\mathbf{D}_{k,1}$ contains the top $n_k$ rows of $\mathbf{D}_k$, and $\mathbf{D}_{k,2}$ contains the remaining $\bar{n}_k-n_k$ rows, since $\mathrm{rank}(\mathbf{D}_k)=\min(n_k,\bar{n}_k)=n_k$. Hence the selection of $\mathbf{D}_{k,2}$ does not affect the optimal precoder design \cite{Tran07}.
It is noted that here we employ the GQRD-base algorithm with a lower computation complexity in \cite{Tran07} to find the optimal precoder given the sum power constraint. The detailed algorithm is however omitted here for brevity as it is not related to the contribution of this work. Interested readers can refer to \cite{Tran07} for details.
In this sense, the rate of user $k$ at epoch $i$ in (\ref{eq:zf_DPC_user_k}) can be transformed to be
\begin{align}
R_k^{ZF-SPC}(i)=\log|\mathbf{I}+\mathbf{L}_k\mathbf{D}_{k,1}(i)\mathbf{D}_{k,1}^H(i)\mathbf{L}_k^H|.
\end{align}
Assigning a weighting factor to user $k$, i.e., $\gamma_k$, the weighted sum throughput hence is given by,
\begin{align}
T=\sum_{i=1}^{N}\sum_{k=1}^{K}\gamma_k l_i
\log|\mathbf{I}+\mathbf{L}_k\mathbf{\Phi}_k(i)\mathbf{L}_k^H|
\end{align}
where $\mathbf{\Phi}_k(i)=\mathbf{D}_{k,1}(i)\mathbf{D}_{k,1}^H(i)$ for short. In the following, we shall aim to find the optimal weighted sum throughput till the deadline $T$ for all considered cases.

\section{Offline Optimization With Ideal Circuit Power Consumption}
The problem to maximize the weighted sum throughput of a MIMO-BC channel with ideal/zero circuit power consumption, termed as {\bf P1}, is formulated as follows,
\begin{align}
\max_{p_i^{sc},p_i^{b},E_i^{sc},E_i^{b}} \quad \sum_{i=1}^{N}\sum_{k=1}^{K}\gamma_kl_i
\log|\mathbf{I}+\mathbf{L}_k\mathbf{\Phi}_k(i)\mathbf{L}_k^H|  \label{eq:opt_ideal}
\end{align}
subject to
\begin{align}
&\sum_{j=1}^{i} p_j^{sc}l_j \le \sum_{j=0}^{i-1}E_j^{sc}, \quad \forall i \label{eq:con_ideal_sc_1} \\
&\sum_{j=0}^{i}E_j^{sc} - \sum_{j=1}^{i} p_j^{sc}l_j \le E_{\max}^{sc}, \quad \forall i \label{eq:con_ideal_sc_2}
\end{align}
\begin{align}
&\sum_{j=1}^i p_j^{b}l_j \le \sum_{j=0}^{i-1} \eta E_j^{b} , \quad \forall i \label{eq:con_ideal_b_1}\\
&\sum_{j=0}^{i} \eta E_j^{b}  - \sum_{j=1}^i p_j^{b}l_j \le  E_{\max}^{b}, \quad \forall i \label{eq:con_ideal_b_2}\\
&p_i^{sc}+p_i^{b } = \sum_{k=1}^{K}\mathbf{\Phi}_k(i) \leq p_{peak}, \quad \forall i. \label{eq:con_ideal_power} \\
& E_i^{sc}+E_i^{b}=E_i  \quad \forall i \label{eq:con_ideal_EH}\\
&E_i^{sc},E_i^{b},p_i^{sc},p_i^{b} \ge 0, \quad \forall i \label{eq:con_ideal_positiveness}
\end{align}
where $\gamma_k$ is the weighting factor associated with the rate of user $k$. (\ref{eq:con_ideal_sc_1}) is the energy causality constraint for SC and (\ref{eq:con_ideal_sc_2}) is the non-energy-overflow constraint of SC. (\ref{eq:con_ideal_b_1}) and (\ref{eq:con_ideal_b_2}) are the energy causality constraint and the  non-energy-overflow constraint at battery, respectively. (\ref{eq:con_ideal_power}) is the physical constraint and the peak power constraint of all antennas for transmission power splitting from SC and battery. In addition, (\ref{eq:con_ideal_EH}) is for the energy harvesting splitting of the hybrid energy storage system and (\ref{eq:con_ideal_positiveness}) is for the positiveness requirement for all design variables.

It is readily observed that {\bf P1} is
a standard convex optimization problem and can be solved
by Karush-Kuhn-Tucker (KKT) conditions.
The Lagrangian function of {\bf P1} is therefore given by,
\begin{align}
&L(E_i^{sc},E_i^{b},p_i^{sc},p_i^b, \Phi_k(i)) \nonumber\\
=&\sum_{i=1}^{N}\sum_{k=1}^{K}\gamma_k l_i
\log|\mathbf{I}+\mathbf{L}_k\mathbf{\Phi}_k(i)\mathbf{L}_k^H| \nonumber\\
& - \sum_{i=1}^N \lambda^{sc}_{1i}\left(\sum_{j=1}^i p_j^{sc}l_j - \sum_{j=0}^{i-1}E_j^{sc} \right) \nonumber\\
& - \sum_{i=1}^{N-1} \lambda^{sc}_{2i} \left( \sum_{j=0}^{i}E_j^{sc} - \sum_{j=1}^i p_j^{sc} l_j - E_{\max}^{sc} \right) \nonumber\\
& -\sum_{i=1}^{N} \lambda^{b}_{1i} \left( \sum_{j=1}^i p_j^{b}l_j - \sum_{j=0}^{i-1} \eta E_j^{b}  \right) \nonumber\\
& -\sum_{i=1}^{N-1} \lambda^{b}_{2i} \left( \sum_{j=0}^{i} \eta E_j^{b}  - \sum_{j=1}^i p_j^{b}l_j - E_{\max}^{b} \right) \nonumber\\
& +\sum_{i=1}^{N} \mu_i \left(p_j^{sc}+p_j^{b } - \sum_{k=1}^{K}\mathbf{\Phi}_k(i) \right) \nonumber\\
& -\sum_{i=1}^{N} \varpi_i \left( \sum_{k=1}^{K}\mathbf{\Phi}_k(i) -p_{peak} \right) \nonumber
\end{align}
\begin{align}
& +\sum_{i=1}^{N} \nu_i \left(E_j^{sc}+E_j^{b}-E_j  \right) + \sum_{i=1}^N\rho_{1i}^{sc}E_i^{sc} \nonumber\\
& + \sum_{i=1}^N\rho_{2i}^{sc}p_i^{sc} + \sum_{i=1}^N\rho_{1i}^{b}E_i^{b}+ \sum_{i=1}^N\rho_{2i}^{b}p_{i}^{b}\label{eq:lagr_ideal}
\end{align}
where $\lambda_{1i}^{sc}$ and $\lambda_{2i}^{sc}$ are the multipliers associated with the energy causality constraint and the energy-non-overflow constraint at epoch $i$ at SC, respectively. $\lambda_{1i}^{b}$ and $\lambda_{2i}^{b}$ are the multipliers associated with the energy causality constraint and the energy-non-overflow constraint at epoch $i$ at battery, respectively. $\mu_i$ and $\varpi_i$ are for the physical constraint of power and the peak power constraint, respectively. $\rho_{1i}^{sc}$ and $\rho_{1i}^{b}$ are for the harvested energy allocated to SC and battery at epoch $i$, respectively. $\rho_{2i}^{sc}$ and $\rho_{2i}^{b}$ are for the transmit power drained from SC and battery at epoch $i$, respectively. %$\rho_{3i}^{sc}$ is for the amount of transferred energy from SC to battery.

By deriving the first-order derivatives of the design variables with respect to the Lagrangian function (\ref{eq:lagr_ideal}), the KKT conditions are given as follows,
\begin{align}
-\sum_{j=i}^N \lambda_{1j}^{sc^*} + \sum_{j=i}^{N-1} \lambda_{2j}^{sc^*}+\mu_i^*+\rho_{2i}^{sc^*}=0  \label{eq:KKT_ideal_p_sc}  \\
%-\sum_{j=i}^N \lambda_{1j}^{sc} + \sum_{j=i}^{N-1} \lambda_{2j}^{sc}+ \sum_{j=i+1}^{N} \eta\lambda_{1j}^b - \sum_{j=i}^{N-1} \eta\lambda_{2j}^b +\rho_{3i}^{sc}=0  \label{eq:KKT_ideal_p_sc_transfer}  \\
-\sum_{j=i}^N \lambda_{1j}^{b^*} + \sum_{j=i}^{N-1} \lambda_{2j}^{b^*}
+ \mu_i^*+\rho_{2i}^{b^*}=0  \label{eq:KKT_ideal_p_b} \\
%\end{align}
%\begin{align}
\sum_{j=i+1}^N \lambda_{1j}^{sc^*} - \sum_{j=i}^{N-1} \lambda_{2j}^{sc^*} + \nu_i^*+\rho_{1i}^{sc^*}=0  \label{eq:KKT_ideal_E_sc}  \\
\sum_{j=i+1}^N \eta \lambda_{1j}^{b^*}-\sum_{j=i}^{N-1} \eta \lambda_{2j}^{b^*} + \nu_i^*+\rho_{1i}^{b^*}=0  \label{eq:KKT_ideal_E_b}  \\
-\left( \varpi_i^*+\mu_i^* \right) \mathbf{I}+ \gamma_k \mathbf{L}_k^H\left(\mathbf{I}+\mathbf{L}_k\mathbf{\Phi}_k^*(i) \mathbf{L}_k^{H}\right)^{-1}\mathbf{L}_k=0 \label{eq:KKT_ideal_precoding}
\end{align}
where the asterisks denote optimality. In addition, (\ref{eq:KKT_ideal_p_sc}) and (\ref{eq:KKT_ideal_p_b}) are KKT conditions for $p^{sc}_i$ and $p^{b}_i$, respectively. (\ref{eq:KKT_ideal_E_sc}) and (\ref{eq:KKT_ideal_E_b}) are KKT conditions for $E^{sc}_i$ and $E^{b}_i$, respectively. (\ref{eq:KKT_ideal_precoding}) is the KKT condition for the optimal precoder design of user $k$ ($k=1,\ldots,K$).
Correspondingly, the slackness conditions are given by,
\begin{align}
&\lambda^{sc^*}_{1i}\left(\sum_{j=1}^i p_j^{sc^*}l_j - \sum_{j=0}^{i-1}E_j^{sc^*} \right) = 0 \label{eq:slack_ideal_p_sc} \\
&\lambda^{sc^*}_{2i} \left( \sum_{j=0}^{i}E_j^{sc^*} - \sum_{j=1}^i p_j^{sc^*} l_j - E_{\max}^{sc} \right) = 0 \label{eq:slack_ideal_Emax}
\end{align}
\begin{align}
&\lambda^{b^*}_{i} \left( \sum_{j=1}^i p_j^{b^*}l_j - \sum_{j=0}^{i-1} \eta E_j^{b^*}  \right)=0 \label{eq:slack_ideal_p_b_1}\\
&\lambda^{b^*}_{2i} \left( \sum_{j=0}^{i} \eta E_j^{b^*}
- \sum_{j=1}^i p_j^{b^*}l_j - E_{\max}^{b} \right)=0
\label{eq:slack_ideal_p_b_2}\\
&\mu_i^* \left(p_j^{sc^*}+p_j^{b^*} - \sum_{k=1}^{K}\mathbf{\Phi}_k^*(i) \right)=0 \label{eq:slack_ideal_power} \\
&\varpi_i^* \left( \sum_{k=1}^{K}\mathbf{\Phi}_k^*(i) -p_{peak} \right)=0
\label{eq:slack_ideal_power_peak} \\
&\nu_i^* \left(E_j^{sc^*}+E_j^{b^*}-E_j  \right)=0 \label{eq:slack_ideal_EH} \\
&\rho_{1i}^{sc^*}E_i^{sc^*} = \rho_{2i}^{sc^*}p_i^{sc^*} =\rho_{1i}^{b^*}E_i^{b^*}=\rho_{2i}^{b^*}p_{i}^{b^*}=0 \label{eq:slack_ideal_positiveness}
\end{align}
%\begin{lemma}\label{lem:power_allocation}
%Firstly, we are interested in the optimal power allocation as well as the optimal precoder design in each epoch.
Combining (\ref{eq:KKT_ideal_p_sc}), (\ref{eq:KKT_ideal_p_b}) as well as
(\ref{eq:KKT_ideal_precoding}) together,
%that
%In the optimal solution to {\bf P1},
the optimal precoder design for the $k$th user is then derived as follows,
\begin{align}
\mathbf{\Phi}_k^*(i) = (\mathbf{L}_k)^{-1}
(\frac{\gamma_k}{\Delta_i^*}\mathbf{L}_k\mathbf{L}_k^H-\mathbf{I})(\mathbf{L}_k^H)^{-1},
\label{eq:power_allocation_precoder}
\end{align}
where
\begin{align}
\Delta_i^*=\mu_i^*+\varpi_i^*&=\sum_{j=i}^N \lambda_{1j}^{sc^*} - \sum_{j=i}^{N-1} \lambda_{2j}^{sc^*}-\rho_{2i}^{sc^*}+\varpi_i^*\nonumber\\
&=\sum_{j=i}^N \lambda_{1j}^{b^*}- \sum_{j=i}^{N-1} \lambda_{2j}^{b^*}-\rho_{2i}^{b^*}+\varpi_i^* \label{eq:sol_ideal_Delta}
\end{align}
Incorporating (\ref{eq:con_ideal_power}) and (\ref{eq:power_allocation_precoder}) and taking some arithmetic operations, the corresponding optimal sum power allocated at the $i$th epoch is given by,
\begin{align}
\sum_{k=1}^K Tr(\mathbf{\Phi}_k^*(i))&=p_{i}^{sc^*} + p_{i}^{b^*} =\sum_{k=1}^K\sum_{\lambda_{ck}}
\left(\frac{\gamma_k}{\Delta_i^*}-\frac{1}{\lambda_{ck}}\right)^+
\label{eq:power_allocation_sum}
\end{align}
where $\lambda_{ck}$ is the nonzero singular values of the matrix $\mathbf{L}_k\mathbf{L}_k^H$ for the $k$th user and $\left( \cdot \right)$ is nonzero if $\{\cdot\}$ is positive and zero otherwise.
%where $1/\lambda_{ck}$ is the water level for the $k$th user at the $i$th epoch for the MIMO-BC channel.
%\end{lemma}
The detailed proof of (\ref{eq:power_allocation_precoder}) and (\ref{eq:power_allocation_sum}) is left to Appendix A and is omitted here for brevity.

Further, noting that $\varpi_i^*$ is positive with active peak power constraint and zero otherwise, the sum power allocated in epoch $i$ can be transformed as follows,
\begin{align}
\sum_{k=1}^K Tr(\mathbf{\Phi}_k^*(i))=\min\left(p_{peak}, \sum_{k=1}^K\sum_{\lambda_{ck}}
\left(\frac{\gamma_k}{\Delta_i^*}-\frac{1}{\lambda_{ck}}\right)^+    \right)
\end{align}
where $\Delta_i^*=\mu_i^*$. In other words, the peak power constraint sets a ceiling level for power allocation and cannot be violated \cite{Palomar}.

In addition to the optimal decoder design of each user as well as the optimal sum power level,
we are also interested in the properties of the optimal solution to {\bf P1}.
Some of them are hence summarized in the following lemmas and propositions.
%\begin{lemma}\label{lem:EH_split}
%In the optimal solution to {\bf P1}, the optimal amount of harvested energy stored in the SC and the battery at the start of $i$th epoch are given by,
%\begin{align}
%&E_i^{sc^*} = \min\left( E_i, E_{\max}-\left( \sum_{j=0}^{i-1}E_j^{sc^*} - \sum_{j=1}^{i}\left( p_j^{sc^*}l_j+\delta_j^*l_j \right)\right)  \right) \label{eq:sol_ideal_EH_split_1}\\
%&E_i^{b^*} = E_i - E_i^{sc^*}\label{eq:sol_ideal_EH_split_1}
%\end{align}
%\end{lemma}

%The detailed proof is simply and is omitted for clarity. Intuitively, due to the energy storage efficiency of the SC, it is optimal to store as many as energy in the SC from the newly arrived harvested energy until its storage capacity ($E_{\max}$) is met and only the residual harvested energy (if available) is stored in the battery.

\begin{lemma}\label{lem:power_allocation_N}
In the optimal solution, the accumulated harvested energy must be used up in the $N$th epoch, i.e.,
\begin{align}
\sum_{i=1}^N p_i^{sc^*}l_i = \sum_{i=0}^{N-1}E_i^{sc^*}  \label{eq:lem_ideal_N_sc}
\end{align}
\begin{align}
\sum_{i=1}^N p_i^{b^*}l_i = \sum_{i=0}^{N-1} \eta E_i^{b^*} \label{eq:lem_ideal_N_b}
\end{align}
and the corresponding slackness conditions
\begin{align}
\lambda_{1N}^{sc^*},\lambda_{1N}^{b^*}>0 \label{eq:lem_ideal_N_slackness}\\
\lambda_{2N}^{sc^*}=\lambda_{2N}^{b^*}=0 \label{eq:lem_ideal_N_slackness_2}
\end{align}
\end{lemma}
The proof is left to Appendix B and is omitted here. It intuitively reveals that the harvested energy should be used up for optimality.

\begin{lemma}\label{lem:sc_lagr_1_2}
In the optimal solution to {\bf P1}, we have
\begin{align}
\lambda_{1i}^{sc^*}\lambda_{2i}^{sc^*}&=0\label{eq:sc_power_property_sc}\\
\lambda_{1i}^{b^*}\lambda_{2i}^{b^*}&=0 \label{eq:sc_power_property_b}
\end{align}
\end{lemma}
\begin{IEEEproof}
Intuitively, it is impossible to meet both (\ref{eq:con_ideal_sc_1}) and (\ref{eq:con_ideal_sc_2}) with equality since $E_{\max}>0$, and therefore at least one of the inequalities is inactively met. From (\ref{eq:slack_ideal_p_sc}) and (\ref{eq:slack_ideal_Emax}), it is concluded that $\lambda_{1i}^{sc^*}$ and/or $\lambda_{2i}^{sc^*}=0$ and (\ref{eq:sc_power_property_sc}) is verified. Similarly, (\ref{eq:sc_power_property_b}) can be demonstrated and the details are omitted here.
\end{IEEEproof}

Intuitively, Lemma \ref{lem:sc_lagr_1_2} reveals that the energy causality constraint and energy-non-overflow constraint cannot be active simultaneously at battery and SC.

\subsection{Monotonic Properties of Transmit Power Transition}
Based on the derived KKT conditions and slackness conditions, some observations on monotonic properties of the sum transmit power allocations are summarized in Lemmas \ref{lem:constant_power_sc}-\ref{lem:nonincreasing_sc} as follows.

\begin{lemma}\label{lem:constant_power_sc}
In the optimal solution to {\bf P1}, if $p_i^{sc^*}$ and $p_{i+1}^{sc^*}>0$ ($p_{i}^{b^*}$ and $p_{i+1}^{b^*}>0$), the associated energy causality constraint and the energy-non-overflow constraint in (\ref{eq:con_ideal_sc_1}) and (\ref{eq:con_ideal_sc_2})  for $p^{sc^*}_i$ (for $p^{b^*}_i$ (\ref{eq:con_ideal_b_1}) and (\ref{eq:con_ideal_b_2}) ) at the $i$th epoch at SC (battery) are inactively met, and the peak power constraint is not actively met in these two epochs, the sum power allocated from the $i$th epoch to the ($i+1$)th epoch remains constant, i.e., $$p_i^{sc^*}+ p_i^{b^*} = p_{i+1}^{sc^*} + p_{i+1}^{b^*}.$$
\end{lemma}
\begin{IEEEproof}
We shall first consider the case of the power drained from the SC.
Since $p_i^{sc^*},p_{i+1}^{sc^*}>0$, we have $\rho_{2i}^{sc^*}=\rho_{2,i+1}^{sc^*}=0$ from the slackness condition in (\ref{eq:slack_ideal_positiveness}). Since (\ref{eq:con_ideal_sc_1}) and (\ref{eq:con_ideal_sc_2}) are inactively met at SC at epoch $i$, we have $\lambda_{1i}^{sc^*}= \lambda_{2i}^{sc^*}=0$ from the slackness conditions in (\ref{eq:slack_ideal_p_sc}) and (\ref{eq:slack_ideal_Emax}). In addition, since the peak power constraint is not actively met, we have $\varpi_i=\varpi_{i+1}=0$.
Combining them together with (\ref{eq:sol_ideal_Delta}), we have
\begin{align}
\Delta_i^*-\Delta_{i+1}^*&=\lambda_{1i}^{sc^*} - \lambda_{2i}^{sc^*}=0
\end{align}
In addition, the proof for the case of the power drained from the battery is identical to that of the power drained from the SC and is omitted here for brevity.
Lemma \ref{lem:constant_power_sc} is verified.
\end{IEEEproof}

%\begin{lemma}\label{lem:constant_power_b}
%In the optimal solution to {\bf P1}, if $p_{i}^{b^*}, p_{i+1}^{b^*}>0$, the energy causality constraint and the energy-non-overflow constraint in (\ref{eq:con_ideal_b_1}) and (\ref{eq:con_ideal_b_2}) at the $i$th epoch at battery are inactively met, and the peak power constraint is inactively met in these two epochs, the sum power allocated from the $i$th epoch to the ($i+1$)th epoch remains constant, i.e., $$p_i^{sc^*}+ p_i^{b^*} = p_{i+1}^{sc^*} + p_{i+1}^{b^*}.$$
%\end{lemma}
%The proof is similar to that of Lemma \ref{lem:constant_power_sc} and is omitted here for brevity.
%\begin{IEEEproof}
%Since $p_{i}^{b^*},p_{i+1}^{b^*}>0$, we have $\rho_{2i}^{b^*}=\rho_{2,i+1}^{b^*}=0$ due to the slackness condition. From the inactive inequality in (\ref{eq:con_ideal_b}) at the $i$th epoch, we deduce that $\lambda_{i}^{b^*}=0$. Combing these together with (\ref{eq:sol_ideal_Delta}) derives,
%\begin{align}
%\Delta_i^*-\Delta_{i+1}^*=\lambda_{i}^{b^*}-\rho_{2i}^{b^*}+\rho_{2,i+1}^{b^*}=0
%\end{align}
%Lemma \ref{lem:constant_power_b} is verified.
%\end{IEEEproof}

\begin{lemma} \label{lem:nondecreasing_sc}
In the optimal solution to {\bf P1}, if $p_i^{sc^*}>0$ or $p_i^{b^*} > 0$, the associated energy causality constraint in (\ref{eq:con_ideal_sc_1}) for $p_i^{sc^*}$ or  (\ref{eq:con_ideal_b_1}) for $p_i^{b^*}$ at the $i$th epoch is actively met, and the peak power constraint is inactively met in the two consecutive epochs $i$ and $i+1$,
the sum power allocated from the $i$th epoch to the ($i+1$)th epoch is strictly increasing, i.e., $$p_i^{sc^*}+ p_i^{b^*} < p_{i+1}^{sc^*} + p_{i+1}^{b^*}.$$
\end{lemma}
\begin{IEEEproof}
We focus on the case at SC as the case at battery is similar and omitted. Since $p_i^{sc^*}>0$, we have $\rho_{2i}^{sc^*}=0$ due to the corresponding slackness condition in (\ref{eq:slack_ideal_positiveness}). In addition, we have $\varpi_i=\varpi_{i+1}=0$ due to the inactive peak power constraints in the two epochs.
Further, as the harvested energy at the SC is used up at the $i$th epoch, we have $\lambda_{1i}^{sc^*}>0$ and $\lambda_{2i}^{sc^*}=0$.
Combining these together derives,
\begin{align}
\Delta_i^*-\Delta_{i+1}^*&=\lambda_{1i}^{sc^*} + \rho_{2,i+1}^{sc^*}>0.
\end{align}
Hence $Tr(\mathbf{\Phi}_k^*(i)) < \sum_{k=1}^K Tr(\mathbf{\Phi}_k^*(i+1))$ and Lemma \ref{lem:nondecreasing_sc} is verified.
\end{IEEEproof}

%\begin{lemma}\label{lem:nondecreasing_b}
%If $p_i^{b^*} \neq 0$ in the optimal solution to {\bf P1},
%the inequality in (\ref{eq:con_ideal_b_1}) at the $i$th epoch is actively met, and the peak power constraint is inactively met in the two consecutive epochs $i$ and $i+1$,
%the sum power allocated from both the SC and the battery is strictly increasing from the $i$th epoch to the $(i+1)$th epoch.
%\end{lemma}
%The proof is similar to that of Lemma \ref{lem:nondecreasing_sc} and is omitted here.

%\textit{Remark}: It is further concluded that the sum power is strictly increasing if $p_i^{b^*}=0$ and/or the energy causality constraint at the battery for the $i$th epoch is active. i.e., $\lambda_i^{b^*}>0$.

\begin{lemma} \label{lem:nonincreasing_sc}
In the optimal solution to {\bf P1}, if $p_{i+1}^{sc^*}>0$ or $p_{i+1}^{b^*} > 0$, the associated energy-non-overflow constraint in (\ref{eq:con_ideal_sc_2}) or (\ref{eq:con_ideal_b_2}) at the $i$th epoch is actively met, and the peak power constraint is inactively met in epoch $i$ and $i+1$, the sum power allocated from the $i$th epoch to the ($i+1$)th epoch is decreasing, i.e., $$p_i^{sc^*}+ p_i^{b^*} > p_{i+1}^{sc^*} + p_{i+1}^{b^*}.$$
\end{lemma}
\begin{IEEEproof}
Since $p_{i+1}^{sc^*}>0$, we have $\rho_{2,i+1}^{sc^*}=0$ due to the corresponding slackness condition in (\ref{eq:slack_ideal_positiveness}). In addition, we have $\varpi_i=\varpi_{i+1}=0$ due to the inactive peak power constraints in the two epochs.
Further, the energy-non-overflow constraint is actively met at the $i$th epoch at SC, we have $\lambda_{1i}^{sc^*}=0$ and $\lambda_{2i}^{sc^*}>0$.
Combining these together derives,
\begin{align}
\Delta_i^*-\Delta_{i+1}^*&=-\lambda_{2i}^{sc^*} - \rho_{2i}^{sc^*}<0.
\end{align}
Note that the proof of the case for the power drained from the battery is similar and is omitted.
Lemma \ref{lem:nonincreasing_sc} is hence proved.
\end{IEEEproof}

\begin{rem} Lemmas \ref{lem:constant_power_sc}-\ref{lem:nonincreasing_sc} disclose some general monotonic properties of the optimal power allocated from the SC and the battery through the epochs, where the exact expressions for the optimal precoder design and sum power allocations are however presented in (\ref{eq:power_allocation_precoder}) and (\ref{eq:power_allocation_sum}).
\end{rem}

\section{Optimization With Maximal Circuit Power Consumption}
In this section, we investigate the case with a constant circuit power consumption in data transmission, where the circuit is assumed to operate with the maximum power consumption. It hence can serve as a lower bound of the performance under the general circuit power consumption model. Specifically, the maximum circuit power consumption is denoted by $\epsilon$. It is also assumed that the circuit operates only when the base station is in the active transmission state. Therefore, the total power assumption is given by $p_{sc}(t)+p_b(t) + \epsilon I_{\{p_{sc}(t)+p_b(t)>0\}}$, where $I_{\{\cdot\}}$ is an indication function which returns unity if the event $\{\cdot\}$ is true and zero otherwise. In the following, we shall first discuss the optimization problem over a single epoch to gain some insights and then naturally extend to the scenario of multiple epochs.

\subsection{The Scenario of A Single Epoch}
To be intuitive, we start by considering transmission over a single epoch. Assuming that $E^{sc}$ and $E^{b}$ are available at the SC and the battery before the start of transmission, respectively, the weighted sum throughput optimization problem over a single-epoch, termed as {\bf P2}, can be formulated as follows.
\begin{align}
\max_{p^{sc}(t),p^b(t),\epsilon^{sc}(t),\epsilon^{b}(t)} \quad \int_{0}^t\sum_{k=1}^{K}\gamma_k
\log|\mathbf{I}+\mathbf{L}_k\mathbf{\Phi}_k(t)\mathbf{L}_k^H|  \label{eq:opt_ideal}
\end{align}
subject to
\begin{align}
&\int_{0}^t p^{sc}(t)+\epsilon^{sc}(t) \leq E^{sc} \\
&\int_0^t p^{b}(t)+\epsilon^{b}(t)   \leq \eta E^{b}  \\
&\epsilon^{sc}(t)+\epsilon^{b}(t)=\epsilon I_{\{p^{sc}(t)+p^b(t)>0\}} \\
&p^{sc}(t)+p^{b}(t) = \sum_{k=1}^K Tr(\mathbf{\Phi}_k(t))\leq p_{peak}
\end{align}

Due to the concavity of the logarithm function, $p^{sc}(t)+p_b(t)$ must remain constant whenever the total transmit power is positive. It is also noted that, unlike \cite{Xu12}, {\bf P2'} has an infinite number of optimal solutions since one can arbitrarily drain energy from the SC and the battery as long as $p^*(t)=p^{sc^*}(t)+p^{b^*}(t)$ is satisfied (where asterisks denote optimality).
Therefore, to simplify {\bf P2}, by assuming that only $0<\tau \leq t$ of $t$ is used for transmission, the following problem, termed as {\bf P2'}, can be formulated by,
\begin{align}
\max_{\tau,p^{sc},p_b,\epsilon^{sc},\epsilon^{b}} \quad \tau\sum_{k=1}^{K}\gamma_k
\log|\mathbf{I}+\mathbf{L}_k\mathbf{\Phi}_k\mathbf{L}_k^H|  \label{eq:opt_cost_P2'}
\end{align}
subject to
\begin{align}
&\tau(p^{sc}+\epsilon^{sc}) \leq E^{sc} \label{eq:con_cost_P2'_sc}\\
&\tau(p^{b}+\epsilon^{b})   \leq \eta E^{b}  \label{eq:con_cost_P2'_b}\\
&\epsilon^{sc}+\epsilon^{b}=\epsilon  \label{eq:con_cost_P2'_cost} \\
&p^{sc}+p^{b} = \sum_{k=1}^K Tr(\Phi_k) \leq p_{peak} \label{eq:con_cost_P2'_powerequality}\\
&0 < \tau \leq t \label{eq:con_cost_P2'_physical}
\end{align}
It is readily found that, to achieve optimality, (\ref{eq:con_cost_P2'_sc}) and (\ref{eq:con_cost_P2'_b}) must hold in equality. Hence, we have
\begin{align}
\tau = \frac{E^{sc}}{p^{sc}+\epsilon^{sc}}=\frac{\eta E^b}{p^{b}+\epsilon^{b}}
=\frac{E^{sc}+\eta E^b}{p^{sc}+p^{b}+\epsilon}
\end{align}

Therefore, relaxing the peak power constraint, the objective function of {\bf P2'} in (\ref{eq:opt_cost_P2'}) can be further simplified as,
\begin{align}
\max_{p^{sc},\,p_b}\,\frac{E^{sc}+\eta E^b}{p^{sc}+p^{b}+\epsilon}\sum_{k=1}^{K}\gamma_k
\log|\mathbf{I}+\mathbf{L}_k\mathbf{\Phi}_k\mathbf{L}_k^H| \label{eq:opt_cost_P2'_1}
\end{align}
with the constraint in (\ref{eq:con_cost_P2'_powerequality}).

It is readily observed that {\bf P2'} indeed aims to optimize the energy usage efficiency given the rate weight of each user.
Incorporating (\ref{eq:con_cost_P2'_powerequality}) into the simplified version of {\bf P2'} results in a non-conditional optimization problem, as shown below,
\begin{align}
\max_{Tr(\mathbf{\Phi}_k)} \frac{E^{sc}+ \eta E^b}{\sum_{k=1}^K Tr(\mathbf{\Phi}_k)+\epsilon}\sum_{k=1}^{K}\gamma_k
\log|\mathbf{I}+\mathbf{L}_k\mathbf{\Phi}_k\mathbf{L}_k^H|
\label{eq:opt_cost_P2'_2}
\end{align}
It can be solved by setting the first order derivative of each $\mathbf{\Phi}_k$ ($k=1,\ldots,M$)
to zero, i.e.,
\begin{align}
&\gamma_k\left( \sum_{k=1}^K Tr(\mathbf{\Phi}_k)+\epsilon \right)\mathbf{L}_k^H\left(\mathbf{I}+\mathbf{L}_k\mathbf{\Phi}_k(i) \mathbf{L}_k^H\right)^{-1}\mathbf{L}_k - \nonumber\\
&\sum_{n=1}^K \gamma_n \log|\mathbf{I}+\mathbf{L}_n\mathbf{\Phi}_n\mathbf{L}_n^H| \cdot \mathbf{I}=0, \quad k=1,\ldots,K. \label{eq:deriv_cost_P2'}
\end{align}

By solving the equations numerically ( the exact expressions cannot be derived as (\ref{eq:deriv_cost_P2'}) is a transcendental equation), the optimal ($\mathbf{\Phi}_k$) ($k=1,\ldots,K$) as well as the sum transmit power $\sum_{k=1}^K Tr(\mathbf{\Phi}_k)$ are therefore determined.

Assuming $p^o=\sum_{k=1}^K Tr(\mathbf{\Phi}_k^*)$ is the optimal solution to (\ref{eq:opt_cost_P2'_2}), by comparing $\left(E^{sc}+\eta E^b\right)/\left(p^o+\epsilon\right)$ with $t$, as well as taking into account the peak power constraint $p_{peak}$, the optimal sum power to {\bf P2'} is therefore given by,
\begin{itemize}
\item If $p^o < p_{peak}$, we have
\begin{align}
p^{sc^*}+p^{b^*}=&
\left \{
\begin{array}{lll}
p^o, & \mbox{if $E_{tol} < t(p^o+\epsilon)$,}\\
p_{peak}, & \mbox{if $E_{tol} > t(p_{peak}+\epsilon)$,}\\
\frac{E_{tol}}{t}-\epsilon, & \mbox{otherwise.}
\end{array}
\right. \label{eq:opt_power_allocation_P2'}
%\min \{p^o,p_{peak},\frac{E^{sc}+ \eta E^b}{t}-\epsilon \} \label{eq:opt_power_allocation_P2'}
\end{align}
where $E_{tol}=E^{sc}+ \eta E^b$.
\item Otherwise we have
\begin{align}
p^{sc^*}+p^{b^*}=p_{peak}. \label{eq:opt_power_allocation_P2'_2}
\end{align}
\end{itemize}
Note that (\ref{eq:opt_power_allocation_P2'}) follows from that if the available energy cannot support the optimal power level for the entire epoch, we should transmit $p^o$ for optimality, otherwise,
the sum transmit power level is determined by the minimum of the peak power and the power level of $\frac{E_{tol}}{t}-\epsilon$.
For (\ref{eq:opt_power_allocation_P2'_2}), however, the optimal power level $p^o$ is not achievable as it violates the peak power constraint.

In addition, the optimal transmission duration $\tau$ is given by
\begin{align}
\tau^* =& \min \{t,\frac{E^{sc}+ \eta E^b}{p^{sc^*}+p^{b^*}+\epsilon} \}\label{eq:opt_tau_allocation_P2'}
\end{align}
%\begin{align}
%L(P) = & \frac{E^{sc}+E_b}{p^{sc}+p^{b}+\epsilon}\sum_{k=1}^{K}\gamma_k\log|\mathbf{I}+\mathbf{L_k\Phi_kL_k^H}|\nonumber %\\
%&+\mu(p^{sc}+p^{b} - \sum_{k=1}^K Tr(\Phi_k)) \nonumber\\
%&-
%\end{align}
and the optimal circuit consumed power splitting at SC and battery are
\begin{align}
\epsilon^{sc^*}=\frac{E^{sc}-p^{sc^*}\tau^*}{\tau^*}
%\left \{
%\begin{array}{ll}
%\frac{E^{sc}}{E^{sc}+\eta E^b}\left(p^o+\epsilon\right)-p^{sc^*}, & \mbox{if $E^{sc}+ \eta E^b \leq t(p^o+\epsilon)$,}\\
%\frac{E^{sc}}{t}-\epsilon, & \mbox{otherwise.}
%\end{array}
%\right.
\label{eq:opt_cost_allocation_sc_P2'}\\
%\end{align}
%\begin{align}
\epsilon^{b^*}=\frac{E^{b}-p^{b^*}\tau^*}{\tau^*}
%\left \{
%\begin{array}{ll}
%\frac{\eta E^{b}}{E^{sc}+\eta E^b}\left(p^o+\epsilon\right)-p^{b^*}, & \mbox{if $E^{sc}+ \eta E^b \leq t(p^o+\epsilon)$,}\\
%\frac{E^{b}}{t}-\epsilon, & \mbox{otherwise.}
%\end{array}
%\right.
\label{eq:opt_cost_allocation_b_P2'}
\end{align}
%where $\tau^* = (E^{sc}+ \eta E^b) / (p^*+\epsilon) $ if $E^{sc}+ \eta E^b \leq t(p^*+\epsilon)$ and $t$ otherwise.
Note that {\bf P2'} has an infinite number of optimal solutions as $p_{sc}^*$ and $p_{b}^*$ can be arbitrarily drained from SC and battery as long as the constraints in (\ref{eq:con_cost_P2'_sc})-(\ref{eq:con_cost_P2'_physical}) are satisfied.
It is also noted that the derived optimal sum power $p^o$ is
independent of the harvested energies $E^{sc}$ and $E^b$ and
is only related to the non-ideal circuit power consumption.
%Physically, if $\tau<t$, it is observed that the global optimal energy efficiency is obtained. On the other hand, the optimal transmit power across $t$ is a local optimal solution if $\tau>t$ (although it is larger than $p^*$), as the global optimal solution is not attainable in this case.

\begin{figure}[t]
   \centering
   \includegraphics[width = 8cm]{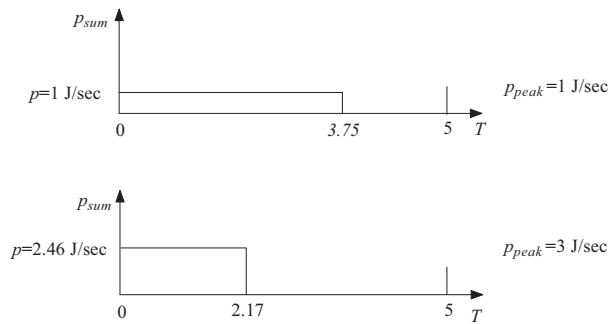}
   \caption{An example of the optimal sum power level and the optimal transmission duration $\tau$ over one epoch versus different peak power constraints, where $p^o=2.46$ J/sec.} \label{fig:peak}
   \end{figure}

Fig. \ref{fig:peak} shows an example of showing the introduction of $p_{peak}$ and the existence of $p^o$ over one epoch, where $E_{\max}^{sc}=5$ J, $E_{\max}^{b}=100$ J, $T=5$ sec, $E^{sc}=5$ J, $E^{b}=5$ J, $\eta = 0.5$ and $\epsilon=1$ J/sec. It is observed that if $p_{peak}<p^o$, the optimal sum power level is always $p_{peak}$. However, if $p_{peak}>p^o$, the optimal sum power level is $p^o$ if $\tau < T$.

\begin{figure}[t]
   \centering
   \includegraphics[width = 7.2cm]{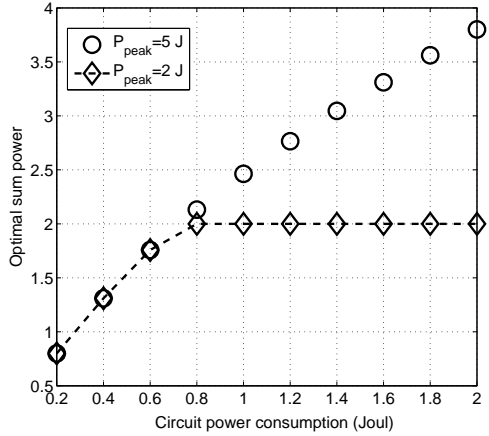}
   \caption{An example of the optimal sum power level and the optimal transmission duration $\tau$ over one epoch versus different peak power constraints.} \label{fig:impact}
   \end{figure}

Further, Fig. \ref{fig:impact} shows the optimal sum power level versus $\epsilon$ subject to the peak power constraint $p_{peak}$, where we set $E_{\max}^{sc}=5$ J, $E_{\max}^{b}=100$ J, $T=5$ sec, $E^{sc}=5$ J, $E^{b}=5$ J, $\eta = 0.6$.
It is observed that the optimal sum power level increases with the increasing of $\epsilon$ till it reaches the peak power constraint and remains at the peak power level, which hence further validates (\ref{eq:opt_power_allocation_P2'}) and (\ref{eq:opt_power_allocation_P2'_2}). The intuition follows from that, with higher $\epsilon$, the BS should transmit at a higher transmit power level and hence spend less time in transmission to compensate for the circuit energy consumption for optimality.


\subsection{The Scenario of Multi-Epoch}
Based on the single epoch optimization problem, for the scenario of $N$ epochs, we can partition each epoch $l_i$ into a transmission interval $\tau_i$ and a silent interval $1-\tau_i$ and the base station transmits during the $\tau_i$ interval and keeps silent in the rest of the $i$th epoch. The associated constraints can be given by,
\begin{align}
&\sum_{j=1}^i \left( p_j^{sc} + \epsilon^{sc}_j \right)\tau_j \le \sum_{j=0}^{i-1}E_j^{sc}, \quad \forall i \label{eq:con_cost_sc_1} \\
&\sum_{j=0}^{i}E_j^{sc} - \sum_{j=1}^i \left( p_j^{sc} + \epsilon^{sc}_j \right)\tau_j \le E_{\max}^{sc}, \quad \forall i \label{eq:con_cost_sc_2} \\
&\sum_{j=1}^i \left(p_j^{b}+\epsilon^{b}_j \right)l_j \le \sum_{j=0}^{i-1} \eta E_j^{b}, \quad \forall i \label{eq:con_cost_b_1} \\
&\sum_{j=0}^{i-1} \eta E_j^{b} - \sum_{j=1}^i \left(p_j^{b} + \epsilon^{b}_j \right)l_j \le E_{\max}^{b}, \quad \forall i \label{eq:con_cost_b_2} \\
&\epsilon_{i}^{sc}+\epsilon_{i}^{b}=\epsilon, \quad \forall i \label{eq:con_cost_cost}\\
&0 \leq \tau_i \leq l_i, \quad \forall i \label{eq:con_cost_time_ratio}\\
&E_i^{sc},E_i^{b},p_i^{sc},p_i^{b}, \epsilon^{sc}_i,\epsilon^{b}_i \ge 0, \quad \forall i \label{eq:con_cost_positiveness}
\end{align}
and the harvested energy splitting constraint and sum transmit power usage constraint in (\ref{eq:con_ideal_power}) and (\ref{eq:con_ideal_EH}), respectively.

Therefore, the problem to maximize the weighted sum throughput of a MIMO-BC channel with the non-ideal circuit power consumption over multiple epochs, termed as {\bf P3}, can be formulated as follows,
\begin{align}
\max_{ p_{i}^{sc},p_i^{b},E_i^{sc},E_i^{b},\epsilon^{sc}_i,\epsilon^{b}_i, \tau_i} \quad \sum_{i=1}^{N}\sum_{k=1}^{K}\gamma_k
\tau_i\log|\mathbf{I}+\mathbf{L}_k\mathbf{\Phi}_k(i)\mathbf{L}_k^H|
\end{align}
subject to the constraints in (\ref{eq:con_ideal_power}), (\ref{eq:con_ideal_EH}) and (\ref{eq:con_cost_sc_1})-(\ref{eq:con_cost_positiveness}).

Note that {\bf P3} is a nonconvex optimization problem, but in nature geometric programming. To circumvent this difficulty, {\bf P3} need to be transformed to be in convex form and some variables are introduced as: $\alpha^{sc}_i=p_i^{sc}\tau_i$, $\sigma^{sc}_i=\epsilon_i^{sc}\tau_i$, $\alpha^{b}_i=p_i^{b}\tau_i$, $\sigma^{b}_i=\epsilon_i^{b}\tau_i$ and $\mathbf{\Phi}_k(i)\tau_i=\mathbf{\Theta}_k(i)$. The constraints are therefore given as follows,
\begin{align}
&\sum_{j=1}^i\left( \alpha_j^{sc}+ \sigma^{sc}_j \right)\le \sum_{j=0}^{i-1}E_j^{sc}, \quad \forall i \label{eq:con_cost_sc_1_new} \\
&\sum_{j=0}^{i}E_j^{sc} - \sum_{j=1}^i \left(\alpha_j^{sc}+ \sigma^{sc}_j \right) \le E_{\max}^{sc}, \quad \forall i \label{eq:con_cost_sc_2_new} \\
&\sum_{j=1}^i\left( \alpha_j^{b}+ \sigma^{b}_j \right)\le \sum_{j=0}^{i-1} \eta E_j^{b} , \quad \forall i \label{eq:con_cost_b_1_new} \\
&\sum_{j=0}^{i-1} \eta E_j^{b}  - \sum_{j=1}^i\left( \alpha_j^{b}+ \sigma^{b}_j \right) \le E_{\max}^{b}, \quad \forall i \label{eq:con_cost_b_1_new} \\
&\sigma^{sc}_i+\sigma^{b}_i=\epsilon \tau_i, \quad \forall i \label{eq:con_cost_cost_new} \\
&\alpha_i^{sc}+\alpha_i^{b} = \sum_{k=1}^K \mathbf{\Theta}_k(i), \quad \forall i\label{eq:con_cost_power_new} \\
&\sum_{k=1}^K \mathbf{\Theta}_k(i) \leq  \tau_ip_{peak} , \quad \forall i  \label{eq:con_cost_power_new_peak}
\end{align}
The objective function in {\bf P3} is correspondingly transformed to be
\begin{align}
\max_{ \alpha_{i}^{sc},\alpha_{i}^{b},E_i^{sc},E_i^{b},\sigma_i^{sc},\sigma_i^{b},\tau_i} \quad \sum_{i=1}^{N}\sum_{k=1}^{K}\gamma_k\tau_i
\log|\mathbf{I}+
\mathbf{L}_k\frac{\mathbf{\Theta}_k(i)}{\tau_i}\mathbf{L}_k^H|. \label{eq:obj_cost_convex_form}
\end{align}

Following from the convex preservation property of the perspective operation of convex functions, (\ref{eq:obj_cost_convex_form}) is convex with respect to $\tau_i$ and the transformed version of {\bf P3} is a convex optimization problem as the constraints are affine functions with respect to the design parameters.
The associated Lagrangian function is however omitted for brevity and
the KKT conditions are thereby given as follows,
\begin{align}
&-\sum_{j=i}^N \lambda_{1j}^{sc^*} + \sum_{j=i}^{N-1} \lambda_{2j}^{sc^*}+\mu_i^*+\rho_{2i}^{sc^*}=0  \label{eq:KKT_cost_p_sc}  \\
%-\sum_{j=i}^N \lambda_{1j}^{sc} + \sum_{j=i}^{N-1} \lambda_{2j}^{sc}+ \sum_{j=i+1}^{N} \eta\Theta\lambda_i^b +\rho_{3i}^{sc}=0  \label{eq:KKT_cost_p_sc_transfer}  \\
&-\sum_{j=i}^N \lambda_{1j}^{sc^*}
+ \sum_{j=i}^{N-1} \lambda_{2j}^{sc^*}
+\omega_i^*+\rho_{3i}^{sc^*}=0  \label{eq:KKT_cost_cost_sc}  \\
%\end{align}
%\begin{align}
&-\sum_{j=i}^N \lambda_{1j}^{b^*} + \sum_{j=i}^{N-1} \lambda_{2j}^{b^*} + \mu_i^*+\rho_{2i}^{b^*}=0  \label{eq:KKT_cost_p_b} \\
&-\sum_{j=i}^N \lambda_{1j}^{b^*} + \sum_{j=i}^{N-1} \lambda_{2j}^{b^*} + \omega_i^*+\rho_{3i}^{b^*}=0  \label{eq:KKT_cost_cost_b} \\
&\sum_{j=i+1}^N \lambda_{1j}^{sc^*} - \sum_{j=i}^{N-1}
\lambda_{2j}^{sc^*} + \nu_i^*+\rho_{1i}^{sc^*}=0
\label{eq:KKT_cost_E_sc}
\end{align}
\begin{align}
&\sum_{j=i+1}^N \eta \lambda_{1j}^{b^*} - \sum_{j=i}^{N-1}
\lambda_{2j}^{b^*} + \nu_i^* + \rho_{1i}^{b^*}=0  \label{eq:KKT_cost_E_b}  \\
&-\left(\varpi_i^*+\mu_i\right)\mathbf{I}+ \frac{\gamma_k}{\tau_i^*} \mathbf{L_k^H}\left(\mathbf{I}+\mathbf{L_k}\frac{\mathbf{\Theta_k^*}(i)}{\tau_i^*} \mathbf{L_k^H}\right)^{-1}\mathbf{L_k}=0 \label{eq:KKT_cost_precoding}\\
&-\sum_{k=1}^{K}\frac{\gamma_k}{\tau_i^*}
Tr\left(\left(\mathbf{I}+\mathbf{L}_k\frac{\mathbf{\Theta}_k^*(i)}{\tau_i^*} \mathbf{L}_k^H\right)^{-1}\mathbf{L}_k\mathbf{\Theta}_k^*(i) \mathbf{L}_k^H\right) \nonumber \\
&+ \sum_{k=1}^{K}\gamma_k
\log|\mathbf{I}+\mathbf{L}_k\frac{\mathbf{\Theta}_k^*(i)}{\tau_i^*}
\mathbf{L}_k^H|+\varpi_i^*p_{peak}  \nonumber \\
&-\epsilon \omega_i^*+\kappa_i^*-z_i^*=0 \label{eq:KKT_cost_timeratio}
\end{align}

The associated complementary slackness conditions are given by,
\begin{align}
 &\lambda^{sc^*}_{1i}\left(\sum_{j=1}^i\left( \alpha_j^{sc^*}+ \sigma^{sc^*}_j  \right) - \sum_{j=0}^{i-1}E_j^{sc^*} \right)=0 \\
 &\lambda^{sc^*}_{2i} \left( \sum_{j=0}^{i}E_j^{sc^*} - \sum_{j=1}^i \left(\alpha_j^{sc^*}+ \sigma^{sc^*}_j \right) - E_{\max}^{sc} \right)=0 \\
 &\lambda^{b^*}_{1i} \left( \sum_{j=1}^i \left(\alpha_{j}^{b^*}+ \sigma^{b^*}_j\right) - \sum_{j=0}^{i-1} \eta E_j^{b^*} \right) =0 \\
 &\lambda^{b^*}_{2i} \left( \sum_{j=0}^{i} \eta E_j^{b^*} - \sum_{j=1}^i \left(\alpha_j^{b^*}+ \sigma^{b^*}_j\right)  - E_{\max}^{b} \right)=0\\
 &\mu_i^* \left(\alpha_i^{sc^*}+\alpha_i^{b^*} - \sum_{k=1}^K \mathbf{\Theta}_k^*(i) \right) =0\\
 &\varpi_i^* \left( \sum_{k=1}^K \mathbf{\Theta}_k^*(i) -\tau_i^* p_{peak} \right) =0 \\
 &\nu_i^* \left(E_j^{sc^*}+E_j^{b^*}-E_j  \right)=0 \\
 &\omega_i^*\left(\sigma^{sc^*}_i+\sigma^{b^*}_i-\epsilon \tau_i^*\right)=0\\
 &z_i^*(\tau_i^*-l_i)=0 \\
 &\rho_{1i}^{sc^*}E_i^{sc^*}=\rho_{2i}^{sc^*}\alpha_i^{sc^*} =\rho_{3i}^{sc^*}\sigma_i^{sc^*} =0 \\
 &\rho_{1i}^{b^*}E_i^{b^*} =\rho_{2i}^{b^*}\alpha_{i}^{b^*}=\rho_{3i}^{b^*}\sigma_i^{b^*}=\kappa_i^* \tau_i^* =0
\end{align}
where the asterisk denotes optimality. $\lambda_{1i}^{sc^*}$ and $\lambda_{2i}^{sc^*}$ are the optimal
multipliers associated with the energy causality constraint
and the energy-non-overflow constraint at epoch $i$ at SC,
respectively. $\lambda_{1i}^{b^*}$ and $\lambda_{2i}^{b^*}$ are
the multipliers associated with the energy causality constraint
and the energy-non-overflow constraint at epoch $i$ at battery,
respectively. $\mu_i^*$ and $\varpi_i^*$ are for the physical
constraint of power and the peak power constraint, respectively.
$\nu_i^*$ is for the physical energy splitting constraint at SC and battery and $\omega_i^*$ is for physical circuit power consumption splitting constraint at SC and battery.
$\rho_{1i}^{sc^*}$ and $\rho_{1i}^{b^*}$ are for the positiveness constraints of the harvested energy
allocated to SC and battery at epoch $i$, respectively.
$\rho_{2i}^{sc^*}$ and $\rho_{2i}^{b^*}$ are for the positiveness constraints of the transmit
power drained from SC and battery at epoch $i$, respectively.
$\rho_{3i}^{sc^*}$ and $\rho_{3i}^{b^*}$ are for the positiveness constraints of the amount of energy drained from SC and battery for circuit power consumption, respectively. $z_i^*$ and $\kappa_i^*$ are for the physical constraints of $\tau_i$ at epoch $i$.

Giving the above KKT conditions and slackness conditions, %the optimal solution to {\bf P1}
%\begin{lemma}\label{lem:power_allocation_cost}
the optimal precoder design for the $k$th user of {\bf P2}
can be derived in a similar way for {\bf P1} and is given by,
\begin{align}
\mathbf{\Phi}_k^*(i) = (\mathbf{L}_k)^{-1}(\frac{\gamma_k}{\tau_i^*\Delta_i^*}
\mathbf{L}_k\mathbf{L}_k^H-\mathbf{I})(\mathbf{L}_k^H)^{-1},
\label{eq:power_allocation_precoder_cost}
\end{align}
and the corresponding optimal sum power allocated at the $i$th epoch is
\begin{align}
\sum_{k=1}^K Tr(\mathbf{\Phi}_k^*(i))&=p_{i}^{sc^*} + p_{i}^{b^*} =\sum_{k=1}^K\sum_{\lambda_{ck}}
(\frac{\gamma_k}{\tau_i^*\Delta_i^*}-\frac{1}{\lambda_{ck}})
\end{align}
where
\begin{align}
\Delta_i^*=\mu_i^*+\varpi^*_i&=\sum_{j=i}^N \lambda_{1j}^{sc^*} - \sum_{j=i}^{N-1} \lambda_{2j}^{sc^*}-\rho_{2i}^{sc^*}+\varpi^*_i\nonumber\\
&=\sum_{j=i}^N \lambda_{1j}^{b^*} - \sum_{j=i}^{N-1} \lambda_{2j}^{b^*}-\rho_{2i}^{b^*}+\varpi^*_i \label{eq:solution_cost_Delta}
\end{align}
and $\gamma_k/(\tau_i^*\Delta_i^*)$ can be regarded as the water level for the $k$th user at the $i$th epoch for the MIMO-BC channel.
%\end{lemma}

%The proof is omitted due to its similarity to Lemma \ref{lem:power_allocation}.
To be intuitive, some new observations on the properties of an
optimal solution due to the non-ideal power consumption, are summarized in the following propositions and lemmas.
\begin{Proposition}\label{lem:cost_transfer_timeratio_1}
If $0<\tau_i<l_i$, then $\sum_{k=1}^K Tr(\mathbf{\Phi}_k^*(i))
= \min(p^o,p_{peak})$.
\end{Proposition}
%The proof directly from (\ref{eq:opt_power_allocation_P2'}) for the single epoch case and the details
%are omitted.
\begin{IEEEproof}
%We prove Lemma \ref{lem:cost_transfer_timeratio_1} by contradiction. Firstly, we assume $\sum_{k=1}^K Tr(\mathbf{\Phi}_k^*(i))
%\neq p^*$.
Note that in the $i$th epoch, the energy used for transmission and circuit power consumption is
$$E_i^{allot}=\tau_i^*\left(\sum_{k=1}^K Tr(\mathbf{\Phi}_k^*(i))+\epsilon\right).$$
Hence, a weighted sum throughput optimization problem for the $i$th epoch only with available energy $E_i^{allot}$ is formulated as
\begin{align}
\max_{\mathbf{\Phi}_k(i)} \quad \frac{E_i^{allot}}{\sum_{k=1}^K Tr(\mathbf{\Phi}_k)+\epsilon} \sum_{k=1}^{K}\gamma_k
\log|\mathbf{I}+\mathbf{L}_k\mathbf{\Phi}_k(i)\mathbf{L}_k^H| \label{eq:lem_power_p^*_cost}
\end{align}
and $\mathbf{\Phi}_k^*(i)$ is optimal to the above problem, due to its optimality for the global optimal solution to {\bf P3}.
Note that (\ref{eq:lem_power_p^*_cost}) is exactly (\ref{eq:opt_cost_P2'_2}) for the single epoch optimization problem {\bf P2'} with $E_i^{allot}$ being the amount of available energy.
Hence from (\ref{eq:opt_power_allocation_P2'}) and (\ref{eq:opt_power_allocation_P2'_2}), since $\tau_i <l_i$, $E_i^{allot}$ is used up and  the optimal sum power is $p^o$ if $p^o < p_{peak}$ and $p_{peak}$ otherwise. Proposition \ref{lem:cost_transfer_timeratio_1} is therefore proved.
\end{IEEEproof}

\begin{Proposition}\label{lem:cost_transfer_timeratio_2}
If $\tau_i^*=l_i$, then $\sum_{k=1}^K Tr(\mathbf{\Phi}_k^*(i))
\ge p^o$ if $p^o < p_{peak}$ and otherwise $p_{peak}$.
\end{Proposition}
The proof directly follows from (\ref{eq:opt_power_allocation_P2'}) and (\ref{eq:opt_power_allocation_P2'_2}) for the single epoch case and is omitted.

\begin{lemma}\label{lem:cost_costratio_P3_1}
If $p^{sc^*}_i$ and $p^{b^*}_i$ are both strictly positive and the peak power constraint is inactive in epoch $i$, then $\epsilon^{sc^*}_i,\epsilon^{b^*}_i>0$.
\end{lemma}
\begin{IEEEproof}
If $p^{sc^*}_i,p^{b^*}_i>0$, we have $\rho_{2i}^{sc^*}=\rho_{2i}^{b^*}=0$ from the slackness conditions. In addition, with inactive peak power constraint, we have $\varpi_i^*=0$.
Combining the analysis above with (\ref{eq:solution_cost_Delta}) we arrive at
\begin{align}
\mu_i^*&=\sum_{j=i}^N \lambda_{1j}^{sc^*} - \sum_{j=i}^{N-1} \lambda_{2j}^{sc^*}=\sum_{j=i}^N \lambda_{1j}^{b^*}- \sum_{j=i}^{N-1} \lambda_{2j}^{b^*}. \label{eq:lem_cost_both_positiveness}
\end{align}
Comparing both sides of (\ref{eq:KKT_cost_cost_sc}) and (\ref{eq:KKT_cost_cost_b}) then derives $\rho_{3i}^{sc^*}=\rho_{3i}^{b^*}$ from (\ref{eq:lem_cost_both_positiveness}). Hence there are only two feasible possibilities: 1) $\rho_{3i}^{sc^*}=\rho_{3i}^{b^*}>0$ and 2) $\rho_{3i}^{sc^*}=\rho_{3i}^{b^*}=0$.
If $\rho_{3i}^{sc^*}=\rho_{3i}^{b^*}>0$, we then arrive at $\epsilon^{sc^*}_i=\epsilon^{b^*}_i=0$ due to slackness condition, which however is impossible since $\epsilon^{sc^*}_i+\epsilon^{b^*}_i=\epsilon>0$ with positive sum power at epoch $i$. Therefore, we must have $\rho_{3i}^{sc^*}=\rho_{3i}^{b^*}=0$ and then $\epsilon^{sc^*}_i, \epsilon^{b^*}_i>0$. Lemma \ref{lem:cost_costratio_P3_1} is proved.
\end{IEEEproof}

\begin{lemma}\label{lem:cost_costratio_P3_2}
For the case that $p^{sc^*}_ip^{b^*}_i=0$ and the peak power constraint is inactive in epoch $i$, we have
\begin{itemize}
\item if $p^{sc^*}_i>0$ and $p^{b^*}_i=0$, then $\epsilon^{sc^*}_i>0$ and $\epsilon^{b^*}_i=0$.
\item if $p^{sc^*}_i=0$ and $p^{b^*}_i>0$, then $\epsilon^{sc^*}_i=0$ and $\epsilon^{b^*}_i>0$.
\item if $p^{sc^*}_i=p^{b^*}_i=0$, then $\epsilon^{sc^*}_i=\epsilon^{b^*}_i=0$.
\end{itemize}
\end{lemma}
The proof is omitted due to its similarity to that of Lemma \ref{lem:cost_costratio_P3_1}.

%\begin{lemma}\label{lem:cost_costratio_P3_3}
%If $p^{sc^*}_i=0$ and $p^{b^*}_i>0$, then $\epsilon^{sc^*}_i=0$ and $\epsilon^{b^*}_i>0$.
%\end{lemma}
%The proof is omitted due to its similarity to that of Lemma \ref{lem:cost_costratio_P3_1}.

\begin{rem}
Lemmas \ref{lem:cost_costratio_P3_1}-\ref{lem:cost_costratio_P3_2} reveal some properties of the optimal portion of energy usage for the non-ideal circuit power consumption drained from the SC and the battery.
It is concluded from such lemmas that only two cases are possible for optimality:
1) both energy units contribute a nonzero portion to the transmit power as well as the circuit power consumption. 2) transmit power and the non-ideal circuit power consumption are both drained from only one energy unit (either SC or battery). This observation is useful as discussed in Sec. \ref{Sec:online} for online strategy design.
\end{rem}

In summary, the above propositions and lemmas reveal useful properties on the optimal power allocation in designing a useful algorithm structure for the scenario with a non-ideal circuit power consumption $\epsilon$, as will be presented in the online scheduling part and demonstrated in the numerical part.

\section{Optimization Over The General Circuit Power Consumption Scenario}

In the above sections, the two extreme cases including the ideal ($\epsilon=0$) circuit power consumption and the maximum circuit power consumption are discussed, here we further extend to a more general circuit power consumption model, denoted by $\epsilon(t)=\epsilon(t)I_{\{p_{sc}(t)+p_{b}(t)>0\}}$ ($0 \le \epsilon(t) \le \epsilon$), where $\epsilon(t)$ is assumed to be zero if the transmitter remains silent at $t$. Further, it is assumed to be constant for one epoch during transmission and can change from epoch to epoch. As discovered above, $p^o$ is independent of any other factors except $\epsilon(t)$, and hence we have $p^o(t)=p^o(\epsilon(t))$ as the optimal power level with regard to $\epsilon(t)$, which can be readily determined by solving (\ref{eq:deriv_cost_P2'}) given $\epsilon(t)$. In this sense, one can compute $p^o(t)$ for all possible $\epsilon(t)$ and build a lookup table associating $p^o(t)$ with $\epsilon(t)$. The corresponding optimization problem for this general circuit power consumption model, termed as {\bf P4}, is then formulated as follows.
\begin{align}
\max_{ p_{i}^{sc},p_i^{b},E_i^{sc},E_i^{b},\epsilon^{sc}_i,\epsilon^{b}_i, \tau_i} \quad \sum_{i=1}^{N}\sum_{k=1}^{K}\gamma_k
\tau_i\log|\mathbf{I}+\mathbf{L}_k\mathbf{\Phi}_k(i)\mathbf{L}_k^H|
\end{align}
subject to the constraints in (\ref{eq:con_ideal_power}), (\ref{eq:con_ideal_EH}), (\ref{eq:con_cost_sc_1})-(\ref{eq:con_cost_b_2}), (\ref{eq:con_cost_time_ratio})-(\ref{eq:con_cost_positiveness}),
and the circuit power splitting constraint
\begin{align}
&\epsilon_{i}^{sc}+\epsilon_{i}^{b}=\epsilon(i), \quad \forall i \label{eq:con_cost_cost1}
\end{align}
It is worthy to note that the only difference of {\bf P4} in contrast to {\bf P3} is that $\epsilon$ is replaced by $\epsilon(i)$ in (\ref{eq:con_cost_cost1}).
Akin to {\bf P3}, {\bf P4} can be transformed to be an equivalent convex optimization problem and can be solved efficiently.
Note also that the KKT conditions, the derived optimal power allocations, as well as the Lemmas derived for {\bf P3} are still valid for {\bf P4} (where appropriate $\epsilon$ is replaced by $\epsilon(t)$), and the details are therefore omitted for brevity.

\section{On-Line Strategies}
In this section, some online strategies are considered for ideal/non-ideal circuit power consumptions. Instead of applying dynamic programming, which requires exhaustive search and with exponential complexity, we resort to some simple but intuitive algorithms which can be easily implemented for large-scale applications with promising performance, as discussed in the following.

\subsection{Ideal Circuit Power Consumption With $\epsilon=0$}
A heuristic  time-energy-aware online algorithm for the ideal-circuit case with $\epsilon=0$ is considered here, where for energy efficiency $E_i^{sc^o}$ and $E_b^{sc^o}$ are determined by
\begin{align}
&E_i^{sc^o} = \min \left( E_i, E_{\max}^{sc} -B^{sc}_{i} \right)  \label{eq:sol_ideal_EH_split_1}\\
&E_i^{b^o} = \min \left( E_i - E_i^{sc^o}, E_{\max}^{b}- B^{b}_{i} \right) \label{eq:sol_ideal_EH_split_2}
\end{align}
where $B^{sc}_{i}$ and $B^{b}_{i}$ are the amount of available energy stored at SC and battery at the end of epoch $i$.

In this algorithm, the sum power level of this strategy is determined at the start of each epoch. Specifically, for the $i$th epoch, it is
given by,
%given the available stored energy divided by the remaining time length.
\begin{align}
p^{on}_{i} =\frac{ E^{sc^o}_{i-1}+ B^{sc}_{i-1} + \eta E^{b^o}_{i-1} +\eta B^{b}_{i-1}}{T-\sum_{j=1}^{i-1}l_j}. \label{eq:online_perfect_sum_power}
\end{align}
where the numerator accounts for the totally drainable energy for the $i$th epoch, and the denominator is the remaining time before the deadline.

\subsection{Non-Ideal Circuit Power Consumption With $\epsilon>0$} \label{Sec:online}
Under the case with non-ideal circuit power consumption, it is observed from the optimal solutions to {\bf P2} and {\bf P3} that there exists a threshold power level $\min (p^o,p_{peak})$. If the energy harvested can support a power level less than $\min (p^o,p_{peak})$ over the entire epoch, it is optimal to transmit a portion of the time with the transmit power $\min (p^o,p_{peak})$, otherwise transmit at a constant power (no higher than $p_{peak}$) over the entire epoch. Motivated by such an observation and recalling that $p^o$ is
independent of the harvested energies $E^{sc}$ and $E^b$ and
is only related to $\epsilon$, an online algorithm with promising performance is presented as follows.
\begin{itemize}
%\caption{A Two-Step Iterative Search Algorithm}\label{AWF}
%\SetKwData{Left}{left}\SetKwData{This}{this}\SetKwData{Up}{up}
%\SetKwFunction{Union}{Union}\SetKwFunction{FindCompress}{FindCompress}
%\SetKwInOut{Input}{input}\SetKwInOut{Output}{output}

\item Input: $T$, $E_0^{sc}$, $E_0^{b}$, $p^o$, $p_{peak}$.
\begin{enumerate}
\item Initialization: Given $E_0^{sc}$ and $E_0^{b}$, the sum power level, the optimal precoder design, as well as the transmission duration at the first epoch are determined by (\ref{eq:opt_power_allocation_P2'})-(\ref{eq:opt_power_allocation_P2'_2})
and (\ref{eq:opt_tau_allocation_P2'}) with $E_{tol}=E_{0}^{sc}+\eta E_0^{b}$.
%\begin{align}
%p^{sc^o}_1+p^{b^o}_1=
%\left \{
%\begin{array}{ll}
%p^*, & \mbox{if $E^{sc}_0+ \eta E^b_0 \leq T(p^*+\epsilon)$,}\\
%\frac{E^{sc}_0+\eta E^b_0}{T}-\epsilon, & \mbox{otherwise.}
%\end{array}
%\right. \label{eq:opt_power_allocation_online_cost_1}
%    \end{align}
\item At the start of the $i$th epoch (the $i$th energy arrival), $E_{i-1}^{sc}$ and $E_{i-1}^{b}$ are determined by (\ref{eq:sol_ideal_EH_split_1}) and (\ref{eq:sol_ideal_EH_split_2}). In addition, the sum
power level and the transmission duration of
the $i$th epoch are determined by (\ref{eq:opt_power_allocation_P2'})-(\ref{eq:opt_power_allocation_P2'_2})
and (\ref{eq:opt_tau_allocation_P2'}),
    %\begin{align}
%p^{sc^o}_i+p^{b^o}_i=
%\left \{
%\begin{array}{ll}
%p^*, & \mbox{if $E^{sc}_{i-1}+ E^{sc}_{resi} + \eta E^b_{i-1} +E^{b}_{resi}\leq \left(T-\sum_{j=1}^{i-1}l_j\right)\left(p^*+\epsilon\right)$,}\\
%\frac{E^{sc}_{i-1}+ E^{sc}_{resi} + \eta E^b_{i-1} +E^{b}_{resi}}{T-\sum_{j=1}^{i-1}l_j}-\epsilon, & \mbox{otherwise.}
%\end{array}
%\right. \label{eq:opt_power_allocation_online_cost_2}
    %\end{align}
    where the total available energy hence is
    %\begin{align}
    $E_{tol} = E^{sc}_{i-1}+ B^{sc}_{i-1} + \eta E^b_{i-1} +\eta B^{b}_{i-1}$.%\nonumber
    %E_{resi}^{sc} &= \sum_{j=0}^{i-2}E_j^{sc} - \sum_{j=1}^{i-1} \left(\left( p_j^{sc} + \epsilon^{sc}_j \right)\tau_j +\delta_j l_j  \right), \nonumber \\
    %E_{resi}^{b} &= \sum_{j=0}^{i-2} \eta \left(E_j^{b} %+\delta_j l_j \right) - \sum_{j=1}^{i-1} \left(p_j^{b}+\epsilon^{b}_j \right)l_j. \nonumber
    %\end{align}
\item If $\sum_{j=1}^{i-1}l_j \leq T$, go to Step 2), otherwise terminate.
\end{enumerate}
\item Output: $p^{sc}_i$, $p^{b}_i$, $\epsilon^{sc}_i$, $\epsilon^{b}_i$, $\tau_i$.
\end{itemize}
%Similarly, in the $i$h epoch, the algorithm in \cite{Tran07} is applied in (\ref{eq:power_allocation_precoder_cost}) to find the optimal water-filling level as well as the optimal precoder designs for all users with the determined sum transmit power level.

\begin{rem} It is noted that to improve energy storage efficiency in online scheduling, transmit power is drained from the SC first, i.e., we let
$p_i^{sc}$ be the sum power level and $\epsilon_i^{sc}=\epsilon$ ($p_i^{b}=0$ and $\epsilon_i^{b}=0$) at the start of the transmission duration of the $i$th epoch, and then drain harvested energy from the battery if the energy at SC is used up, i.e., we let $p_i^{b}$ be the sum power level and $\epsilon_i^{b}=\epsilon$.
In this way, we can make full use of SC in the following epochs and the energy storage efficiency as well as the weighted throughput are improved, as observed in Lemma \ref{lem:cost_costratio_P3_2}.
\end{rem}

\begin{rem} The transmitter/base station transmits at the optimal power level $p^o$ (if $p^o < p_{peak}$) if the amount of the harvested energy is not sufficiently large. In this case, it is argued that our algorithm can achieve comparable performance to its offline counterpart,
as will be observed in the numerical results part.
\end{rem}
\begin{rem}
Further, it is noted that for the general $\epsilon(t)$ model,
$p^o$ is replaced by $p^o(\epsilon(t))$ and the above algorithm for the case with a maximum $\epsilon>0$ applies.
\end{rem}
%However, if the amount of harvested energy at each epoch is sufficiently large, there should be some performance penalty due to the uncertainties of the amount of energy harvested and the arrival time of the harvested energy at each epoch. Both cases will be addressed in the numerical part.
%\begin{figure}[t]
%\centering
%\includegraphics[width = 8cm]{one_time_sim_1.eps}
%\caption{Accumulated weighted sum throughput of different policies with time, where $\eta=0.6$. The deadline is $10$ seconds.}
%\label{fig:one_time_sim_1}
%\end{figure}
\begin{figure}[t]
\centering
\includegraphics[width = 8cm]{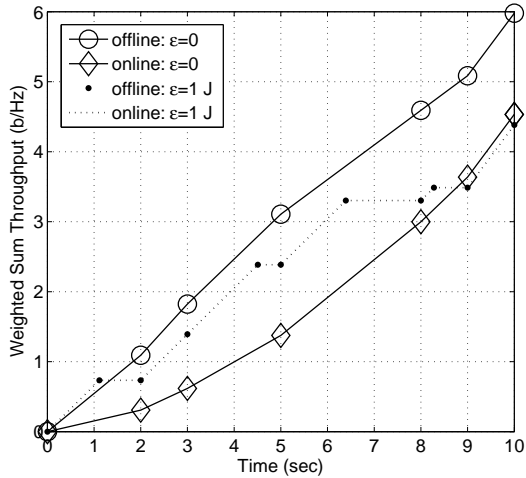}
\caption{Accumulated weighted sum throughput of different policies with time, where $\eta=0.6$. The deadline is $10$ seconds.}
\label{fig:one_time_sim}
\end{figure}
\begin{figure}[t]
\centering
\includegraphics[width = 8cm]{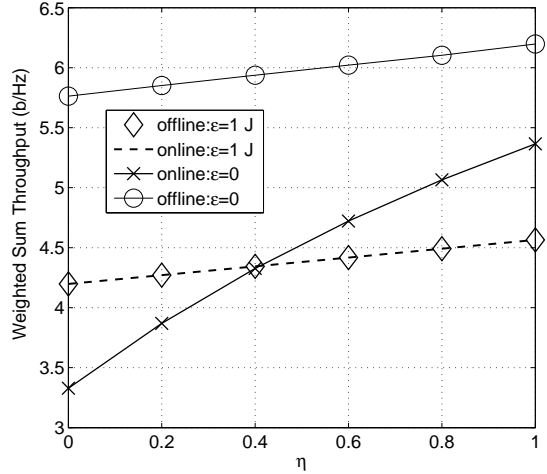}
\caption{Weighted sum throughput of different policies versus varying battery storage efficiency $\eta$. The deadline is $10$ seconds.}
\label{fig:one_time_eta}
\end{figure}
\begin{figure}[t]
\centering
\includegraphics[width = 8cm]{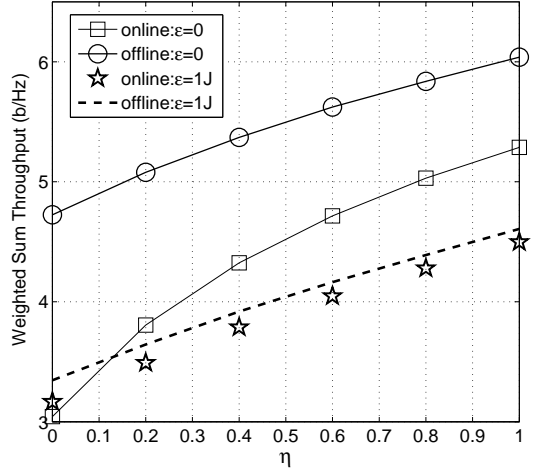}
\caption{Average weighted sum throughput of different policies versus the varying $\eta$, where $E_{avg}=5$ J. The deadline is $10$ seconds.}
\label{fig:avg_eta}
\end{figure}
\begin{figure}[t]
\centering
\includegraphics[width = 8cm]{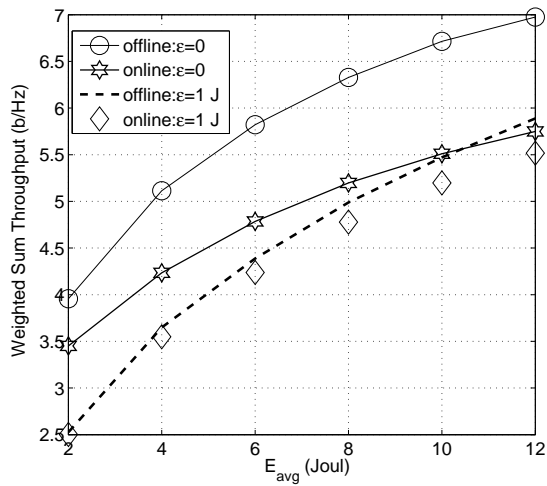}
\caption{Average weighted sum throughput of different policies versus the varying $E_{avg}$, where $\eta=0.6$. The deadline is $10$ seconds.}
\label{fig:arrival_rate}
\end{figure}
\begin{figure}[t]
\centering
\includegraphics[width = 8cm]{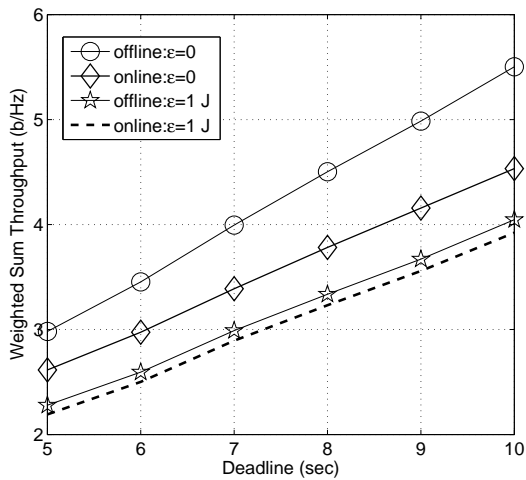}
\caption{Average weighted sum throughput of different policies versus different deadlines, where $\eta=0.6$ and $E_{avg}=5$ J.}
\label{fig:deadline}
\end{figure}
\begin{figure}[t]
\centering
\includegraphics[width = 8cm]{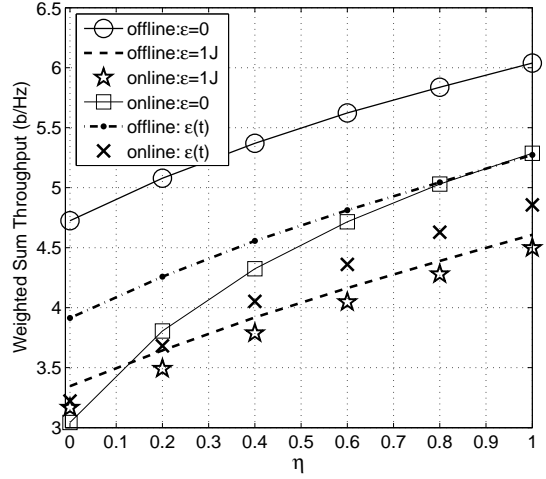}
\caption{Average weighted sum throughput of different policies versus the varying $\eta$, where a general circuit power consumption model is evaluated. The deadline is set to be $10$ seconds.}
\label{fig:avg_eta_random_epsilon}
\end{figure}
\begin{figure}[t]
\centering
\includegraphics[width = 8cm]{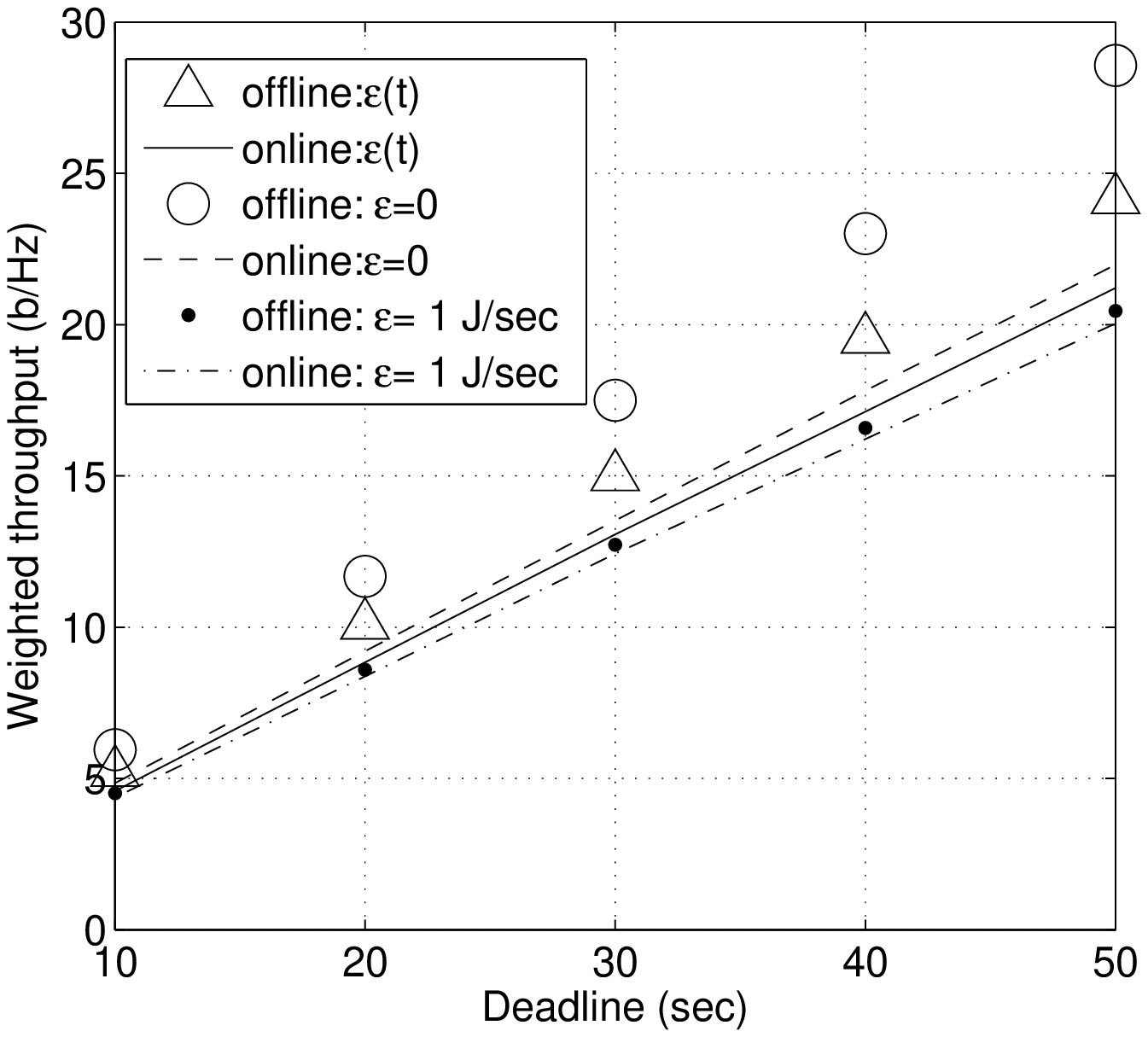}
\caption{Average weighted sum throughput of different policies versus transmission deadline, where a general circuit power consumption model is evaluated. The deadline ranges from $10$ seconds to $50$ seconds and $\eta=0.6$.}
\label{fig:deadline_2}
\end{figure}

\section{Numerical Results}
In this section, both the optimal offline solutions and the heuristic online algorithms are numerically evaluated. The channels between the antennas of all users and those of the transmitter are assumed to be mutually independent Gaussian channels with zero mean and unit variance. The noise is additive white Gaussian noise with zero mean and unity variance. The sum peak power constraint at the transmitter is assumed to be $4$ J/sec. The storage capacities at SC and battery are $5$ J and $100$ J, respectively. The deadline is assumed to last for $10$ seconds, if not otherwise noted.

Firstly, to gain insights, we consider the scenario of deterministic
energy arrivals in Figs. \ref{fig:one_time_sim}-\ref{fig:one_time_eta}. It is assumed that the amounts of energy $E=[4,7,3,5,1,8]$ J arrive at time instants $t=[0,2,3,5,8,9]$ sec.

%In Fig. \ref{fig:one_time_sim}, with the deterministic energy arrival profiles, the weighted sum throughput with respect to time is plotted .
%To demonstrate the benefits of the hybrid energy storage system, the performance of an energy storage system with only the ideal SC unit is also evaluated and compared in Fig. \ref{fig:one_time_sim_1}, given the deterministic energy arrival profiles.
%It is observed that, the weighted sum throughput with a hybrid system including an imperfect battery outperforms that with only the ideal SC storage unit under both the zero/nonzero circuit power consumption scenarios, which further validates the benefits of the battery due to the fact that it can smooth the transmit power across the epochs.

In Fig. \ref{fig:one_time_sim}, the performances with respect to time for both offline and online scheduling are plotted, given the deterministic energy arrival profiles.
It is observed in the ideal circuit power consumption case that, compared with the offline strategy, the online policy performs poorly in the beginning, and their gap shrinks with the reducing remaining time length ($T-\sum_{j=1}^{i-1}l_j$) towards the deadline, as observed in (\ref{eq:online_perfect_sum_power}).
Actually, the online strategy finally captures roughly $75\%$ performance of its offline counterpart in terms of weighted sum throughput.
On the other hand, the performances of both offline and online
strategies match with each other well under the scenario with maximum circuit power consumption.
The advantage of the proposed online policy utilizing the optimal $p^o$ is therefore validated.

In Fig. \ref{fig:one_time_eta}, the performances of different strategies with respect to the varying battery energy efficiency $\eta$ are plotted, given the deterministic energy arrival profiles. It is observed that the offline strategy with ideal circuit power consumption ($\epsilon=0$) outperforms the offline strategy with maximum circuit power consumption ($\epsilon=1$ J/sec), due to the introduction of circuit power consumption. In addition, the online policy for the ideal case with $\epsilon=0$ can capture at least $60\%$ performance compared to its offline counterpart. In fact, the performance gap shrinks with the enhancement of battery storage efficiency. With perfect battery storage efficiency, the online strategy can achieve roughly $86\%$ in terms of the weighted sum throughput.
In addition, for the case with non-ideal circuit power consumption, the performance of the online policy
matches perfectly with that of the offline counterpart, due to
the utilization of the optimal $p^o$.
%as the proposed online policy utilizes the derived $p^o$ and the peak power constraint from the offline part.

%Concretely, for the time-energy adaptive online policy, too much energy is delivered to future slots in the beginning and is not optimal. However, while approaching the deadline, this online policy performs closer to its offline counterpart.

In the following, we consider stochastic energy arrivals. Specifically, we model the energy arrival as a compound Poisson process with uniform density $f_e$ over the interval $[0, 2E_{avg}]$ where $E_{avg}$ is the average harvested amount each time. The arrival rate is taken as $1$ arrival per second. In addition, the amount of the initial energy arrival is assumed to be 5 J at the BS. The weighted throughput are averaged by running $1000$ simulations for accuracy.

In Fig. \ref{fig:avg_eta}, the performances of different strategies with respect to the varying battery energy efficiency $\eta$ are plotted. Similar to the deterministic energy arrival profiles, for the case with zero/ideal circuit power consumption, the online strategy performs better compared to its offline counterpart with the increasing $\eta$, due to the improved energy storage efficiency of battery. Specifically, with $\eta=1$, the online policy captures roughly $86\%$ performance of its offline counterpart, as in the case the SC and the battery can be regarded as one perfect energy storage unit. For the case with maximum/constant circuit power consumption, the online strategy performs quite close to its offline counterpart, i.e., over $92\%$ performance of the offline scheduling can be achieved by employing the proposed online policy, which is promising in practical applications.

In Fig. \ref{fig:arrival_rate} and Fig. \ref{fig:deadline}, the performances of different strategies with respect to the varying $E_{avg}$ and the deadlines are shown, respectively. It is observed in Fig. \ref{fig:arrival_rate} that, with the increasing $E_{avg}$, the performance of all policies monotonically increases. It is also observed in Fig. \ref{fig:deadline} that, under the scenario with zero/ideal circuit power consumption, the time-energy aware online policy performs better with smaller deadlines. In addition, for the case with maximum circuit power consumption, the proposed online policy still performs close to its offline counterpart.

To complete this work, we consider a general circuit power consumption model $\epsilon(t)$ in Fig. \ref{fig:avg_eta_random_epsilon} and Fig. \ref{fig:deadline_2}, where $\epsilon(t)$ is modeled as a stochastic process following the uniform distribution ranging in [$0$, $\epsilon$] at each epoch whenever the sum transmit power is positive.
It is not surprisingly observed in Fig. \ref{fig:avg_eta_random_epsilon} that the performance of the offline scheduling with $\epsilon(t)$ performs worse than the ideal circuit case but outperforms the maximum circuit power consumption case. Interestingly, it is observed that the performance of the online policy with $\epsilon(t)$ can capture no less than $90\%$ performance of its offline counterpart with $\eta \ge 0.4$ and is also promising in practice.

Finally, we evaluate the performance of different schemes with respect to deadline $T$ in Fig. \ref{fig:deadline_2}. It is observed that under ideal circuit power consumption case, the online policy performs worse with the increasing of the deadline, which follows from that saving too much energy for future use absolutely deteriorates performance. Still, it can capture roughly $75\%$ performance of its offline counterpart when $T=50$ sec.
For the case with maximum circuit power consumption, it is still observed that the online policy performs quite close to its offline counterpart. For the case with a random $\epsilon(t)$ circuit power consumption, the gap between the online policy and its offline counterpart slightly enlarges with the increasing of the deadline $T$. However, the online policy can still achieve $88\%$ roughly of its offline counterpart even when $T=50$ sec and is hence promising in practice.

\section{Conclusion}
In this work, a MIMO broadcast channel under the energy harvesting (EH) constraint and the peak power constraint was investigated, where the transmitter was equipped with a hybrid energy storage system consisting of a perfect SC and an inefficient battery. Both elements were assumed to be of limited energy storage capacities.
In addition, two extreme circuitry power consumption cases including the ideal and the maximum (constant) circuit power consumptions were discussed.
The associated optimization problems in offline scheduling were formulated and solved in terms of weighted throughput optimization.
Furthermore, the general case where the circuit power consumption ranges between the two extreme cases was also discussed.
Some intuitive online policies were presented to complement this work. Interestingly, for the maximum circuit power consumption case, a close-to-optimal online policy was presented and its performance was shown to be comparable to its offline counterpart in the numerical results, which is of practical importance in contemporary energy harvesting communication systems with nonnegligible processing power consumption.

Further, it is expected that the circuit power consumption will get lower with the innovative development of circuit technology.
Therefore, how to improve the online scheduling performance by taking such facts into account would be one of the interesting research directions in future. In addition, improving the efficiency of battery storage should also be of fundamental importance in the near future, especially for low-cost sensor networks.

\appendices
\section{Proof of Optimal Precoder Design}
\begin{IEEEproof}
From the KKT condition in (\ref{eq:KKT_ideal_precoding}) for {\bf P1}, we arrive at
\begin{align}
\mathbf{L}_k^H\left(\mathbf{I}+\mathbf{L}_k\mathbf{\Phi}_k(i) \mathbf{L}_k^H\right)^{-1}\mathbf{L}_k=\frac{\Delta_i^*}{\gamma_k}\mathbf{I}.
\label{eq:appendix_inv}
\end{align}
As $\mathbf{L}_k$ is nonsingular, by letting $\mathbf{L}_k=(\mathbf{L}_k^{-1})^{-1}$, $\mathbf{L}_k^H=((\mathbf{L}_k^{H})^{-1})^{-1}$ and by getting the inverse matrix on both sides of (\ref{eq:appendix_inv}), we obtain
\begin{align}
(\mathbf{L}_k^{-1})\left(\mathbf{I}+\mathbf{L}_k\mathbf{\Phi}_k(i) \mathbf{L}_k^H\right)(\mathbf{L}_k^{H})^{-1}=\frac{\gamma_k}{\Delta_i^*}\mathbf{I}.
\end{align}
Applying some arithmetic operations, we hence arrive at
\begin{align}
\mathbf{\Phi}_k(i) =&
\frac{\gamma_k}{\Delta_i^*}\mathbf{I}-(\mathbf{L}_k^{-1})(\mathbf{L}_k^{H})^{-1} \nonumber\\
=&\mathbf{L}_k^{-1}\left(\frac{\gamma_k}{\Delta_i^*}\mathbf{L}_k\mathbf{L}_k^{H}
-\mathbf{I}\right)(\mathbf{L}_k^{H})^{-1}
\end{align}
In addition,
\begin{align}
Tr(\mathbf{\Phi}_k(i))=&Tr(\frac{\gamma_k}{\Delta_i^*}\mathbf{I}-(\mathbf{L}_k^{-1})(\mathbf{L}_k^{H})^{-1} )\nonumber \\
=&Tr(\frac{\gamma_k}{\Delta_i^*}\mathbf{I}
-(\mathbf{L}_k^{H}\mathbf{L}_k)^{-1}) \nonumber\\
=&Tr(\frac{\gamma_k}{\Delta_i^*}\mathbf{I}
-(\mathbf{L}_k\mathbf{L}_k^{H})^{-1}) \label{eq:appendix:trace}\\
=&\sum_{\lambda_{ck}}
(\frac{\gamma_k}{\Delta_i^*}-\frac{1}{\lambda_{ck}})
\end{align}
where $\lambda_{ck}$ is the singular values of the matrix
$\mathbf{L}_k\mathbf{L}_k^{H}$ and (\ref{eq:appendix:trace}) follows from the cyclic property $Tr(\mathbf{AB})=Tr(\mathbf{BA})$ for all matrices $\mathbf{A}$ of dimension $m$ by $n$ and $\mathbf{B}$ of dimension $n$ by $m$.
\end{IEEEproof}

\section{Proof of Lemma \ref{lem:power_allocation_N}}
\begin{IEEEproof}
We prove Lemma \ref{lem:power_allocation_N} by contradiction. Suppose under the optimal solution, the constraints in (\ref{eq:con_ideal_sc_1}) and (\ref{eq:con_ideal_b_1}) are met with strict inequality at the $N$th epoch. In this case, one can always increase the transmit power to meet the corresponding constraint(s) with equality. The weighted sum throughput is therefore improved, which contradicts the optimality. Hence, (\ref{eq:lem_ideal_N_sc}) and (\ref{eq:lem_ideal_N_b}) are verified. From the slackness conditions for the energy causality constraints at SC and battery in (\ref{eq:slack_ideal_p_sc}) and (\ref{eq:slack_ideal_p_b_1}), it is further concluded that $\lambda_{1N}^{sc^*},\lambda_{1N}^{b^*}>0$, and (\ref{eq:lem_ideal_N_slackness}) is verified. Since the energy causality constraint is active at epoch $N$, the non-energy-overflow constraint is hence inactive at the last epoch, and (\ref{eq:lem_ideal_N_slackness}) is immediately deduced.
\end{IEEEproof}
\end{document}